\documentclass[twocolumn]{aastex631}

\usepackage[flushleft]{threeparttable}
\usepackage[caption=false]{subfig}

\shorttitle{MEAD I: ir dust extinction features}
\shortauthors{Decleir et al.}

\graphicspath{{./}{Figures/}}

\begin{document}

\title{A first taste of MEAD (Measuring Extinction and Abundances of Dust) -- I. Diffuse Milky Way interstellar dust extinction features in JWST infrared spectra}

\author[0000-0001-9462-5543]{Marjorie Decleir}
\altaffiliation{ESA Research Fellow}
\affiliation{European Space Agency (ESA), ESA Office, Space Telescope Science Institute (STScI), 3700 San Martin Drive, Baltimore, MD 21218, USA
}

\correspondingauthor{Marjorie Decleir}
\email{mdecleir@stsci.edu}

\author[0000-0001-5340-6774]{Karl D. Gordon}
\affiliation{Space Telescope Science Institute, 3700 San Martin
  Drive, Baltimore, MD 21218, USA}
\affiliation{Sterrenkundig Observatorium, Universiteit Gent,
  Gent, Belgium}

\author{Karl A. Misselt}
\affiliation{Steward Observatory, University of Arizona, 933 N Cherry Ave., Tucson, AZ 85721, USA}

\author[0000-0002-2449-0214]{Burcu G\"unay}
\affiliation{Institut de Ciéncies de l’Espai (ICE-CSIC), Campus UAB, Carrer de Can Magrans S/N, E-08193 Cerdanyola del Vallés, Catalonia, Spain}
\affiliation{Armagh Observatory and Planetarium, Armagh, NI, UK}

\author[0000-0001-6326-7069]{Julia Roman-Duval}
\affiliation{Space Telescope Science Institute, 3700 San Martin
  Drive, Baltimore, MD 21218, USA}

\author[0000-0002-8163-8852]{Sascha T. Zeegers}
\altaffiliation{ESA Research Fellow}
\affiliation{European Space Agency (ESA), European Space Research and Technology Centre (ESTEC), Keplerlaan 1, 2201 AZ Noordwijk, The Netherlands}

\begin{abstract}
We present the initial results of MEAD (Measuring Extinction and Abundances of Dust), with a focus on the dust extinction features observed in our JWST near- and mid-infrared spectra of nine diffuse Milky Way sightlines ($1.2 \leq A(V)\leq 2.5$). For the first time, we find strong correlations between the 10\,\micron\ silicate feature strength and the column densities of Mg, Fe and O in dust. This is consistent with the well-established theory that Mg- and Fe-rich silicates are responsible for this feature. We obtained an average stoichiometry of the silicate grains in our sample of Mg:Fe:O = 1.1:1:11.2, constraining the grain composition. We find variations in the feature properties, indicating that different sightlines contain different types of silicates.  In the average spectrum of our sample, we tentatively detect features around 3.4 and 6.2\,\micron, which are likely caused by aliphatic and aromatic/olefinic hydrocarbons, respectively. If real, to our knowledge, this is the first detection of hydrocarbons in purely diffuse sightlines with $A(V)\leq2.5$, confirming the presence of these grains in diffuse environments. We detected a 3\,\micron\ feature toward HD073882, and tentatively in the sample average, likely caused by water ice (or solid-state water trapped on silicate grains). If confirmed, to our knowledge, this is the first detection of ice in purely diffuse sightlines with $A(V)\leq2.5$, supporting previous findings that these molecules can exist in the diffuse ISM. 
\end{abstract}

\section{Introduction\label{sec:intro}}

Interstellar dust absorbs and scatters a large fraction of star light in the ultraviolet (UV), visible, near- and mid-infrared (NIR/MIR), and re-emits the absorbed energy in the infrared (IR), hence modifying the observed spectral energy distribution of stars, galaxies and other astrophysical objects. In addition, gas and dust are generally well mixed in the interstellar medium (ISM) \citep[e.g.,][]{1978ApJ...224..132B,2022dge..book.....W}, which makes dust a key tracer of the ISM, both locally \citep[e.g.,][]{2011ApJ...741...12B} and at high redshift \citep[e.g.,][]{2014MNRAS.441.1017R}. Furthermore, interstellar dust is fundamental in the processes of star formation and galaxy evolution: dust grains shield the interiors of clouds from the radiation of young stars, allowing molecules (such as molecular hydrogen) to form and exist, the interstellar gas to cool and condense, and new generations of stars to form. Understanding the interstellar dust grain properties is thus crucial to explain star formation and hence galaxy evolution, to better understand the ISM in general, as well as to account for the effects of dust in a range of other studies. 

Interestingly, it is precisely this effect of the dust on the star light that provides insights into the properties of the grains in the ISM. Dust \textit{extinction}, which is the combination of absorption and scattering out of the line of sight, has a clear imprint on the observed spectrum of a background star. Extinction features in the spectrum provide a direct measurement of the composition of dust grains in the ISM intervening between the star and the observer. In the UV, for example, there is a prominent extinction feature at 2175 \AA, which likely originates from absorption by carbonaceous grains \citep[e.g.,][]{1965ApJ...142.1681S}.
At longer wavelengths, in the NIR and MIR, several other features have been detected. The strongest features around 10 and 20\,\micron\ are caused by Si\sbond O stretching and O\sbond Si\sbond O bending in silicate grains, respectively \citep[e.g.,][]{1970Natur.227..822H, 1974ApJ...192L..15D,2010ARA&A..48...21H}. Other absorption features around 3.4 and 6.2\,\micron\ are attributed to C\sbond H and C\dbond C stretching in aliphatic and olefinic/aromatic hydrocarbons, respectively \citep[e.g.,][]{1991ApJ...371..607S, 1994ApJ...437..683P, 1998A&A...337..261S, 2013ApJ...770...78C}. Finally, an absorption feature has been detected around 3\,\micron\ caused by the O\sbond H stretching mode of bulk H\textsubscript{2}O ice in sightlines that go through denser material \citep[e.g.,][]{1997ApJ...490..729W, 2000ApJ...536..347G, 2015ARA&A..53..541B}.

In addition to studying features in the observed spectra, we can measure an extinction curve, representing the amount of extinction as a function of wavelength. The slope of the extinction curve gives an estimate of the average dust grain size along the line of sight, as smaller grains preferentially extinguish light at shorter wavelengths, while larger grains more equally extinguish light at all wavelengths. The total-to-selective extinction ratio $R(V)$, which is defined as $A(V)/E(B-V)$, probes the slope of an extinction curve, and as such the average dust grain size along the line of sight \citep{2022dge..book.....W}. On average, $R(V)=3.1$ in the Milky Way \citep[see e.g.,][]{1989ApJ...345..245C}, with large variations between different sightlines ($R(V) \sim 2 -5$) \citep{1999PASP..111...63F}. Regions in the Galaxy dominated by
smaller/larger grains are characterized by steeper/shallower curves (larger/smaller $E(B-V)$, and thus smaller/larger $R(V)$-values). Most
variations between different Milky Way extinction curves are found to strongly correlate with $R(V)$ (see \cite{2023ApJ...950...86G} for the most recent FUV--MIR Milky Way $R(V)$-dependent extinction curve, which heavily relied on measurements by \cite{2009ApJ...705.1320G}, \cite{2019ApJ...886..108F}, \cite{2021ApJ...916...33G} and \cite{2022ApJ...930...15D}).

Another, more indirect but quantitative way of uncovering the chemical composition of dust grains is provided by measuring the amount of the elements that make up most of the dust (such as carbon (C), silicon (Si), magnesium (Mg), iron (Fe) and oxygen (O)). The amount of an element in the dust is generally obtained by measuring the amount of that element in the gas phase, and subtracting it from an assumed total (gas+dust) amount.
In other words, we estimate the amount of certain elements in the dust by measuring what is ``missing" from the gas. This is often referred to as a \textit{depletion} measurement \citep[see e.g.,][]{2009ApJ...700.1299J}. 

In the MEAD project, short for Measuring Extinction and Abundances of Dust, we combine UV--MIR extinction measurements with elemental abundance measurements in a sample of diffuse Milky Way sightlines. This enables us to correlate the properties of every extinction feature (such as peak wavelength, strength, and width) to the abundance of elements in dust and their ratios (e.g., Mg/Fe). This poses additional constraints on the detailed chemical composition of the dust grains, and allows us to directly assess the accuracy of existing dust grain models such as \cite{2003ARA&A..41..241D,2003ApJ...598.1017D}, \cite{2004ApJS..152..211Z}, Astrodust \citep{2023ApJ...948...55H} and THEMIS \citep{2013A&A...558A..62J,2017A&A...602A..46J,2024A&A...684A..34Y}. To this aim, we obtained data with the Hubble Space Telescope (HST, PID GO 16285, PI: Decleir) to measure elemental abundances, and the James Webb Space Telescope (JWST, PID GO 2459, PI: Decleir) to measure IR extinction. In addition, we use UV extinction curves for this sample from \cite{2009ApJ...705.1320G}, measured from FUSE (Far Ultraviolet Spectroscopic Explorer) and IUE (International Ultraviolet Explorer) spectra. Finally, as the dust grain properties might depend on the environment (such as gas cloud density) in which they reside, we also investigate any correlations with atomic (\ion{H}{1}). and molecular (H\textsubscript{2}) hydrogen column densities, measured by \cite{2023ApJ...944...33V}.

Important to note is that in MEAD, we focus on the \textit{diffuse} ISM in the Milky Way, as opposed to dense molecular clouds. Different definitions for what is diffuse and what is dense can be found in the literature \citep[see e.g.,][]{2022dge..book.....W}. One way is to use the total V-band extinction $A(V)$ to distinguish between diffuse and dense sightlines. However, a high $A(V)$ can also be measured in sightlines without dense material (e.g., in the diffuse sightline toward the hypergiant Cyg OB2-12, which has $A(V)\simeq10$; \cite{2015ApJ...811..110W}). Here, we will refer to diffuse sightlines as those that do not contain (large amounts of) ice, as e.g. measured by the strength of the 3\,\micron\ water ice feature. However, as will be discussed in more detail in Sec.~\ref{sec:ice}, some diffuse sightlines have been shown to have a weak feature around those wavelengths, which could be caused by small amounts of water ice or trapped solid-state water \citep{2021NatAs...5...78P,2024ApJ...965...48P}. This complicates our distinction between diffuse and dense sightlines. Nevertheless, for the purpose of this work, all MEAD sightlines are considered diffuse.

Previous studies have observed and investigated NIR/MIR extinction features, but many of them focus on a single or a handful of heavily extinguished sightlines, for example toward Cyg OB2-12 ($A(V)\simeq10$) \citep{1990MNRAS.243..400A,1998A&A...337..261S, 2016ApJ...830...71F,2020ApJ...895...38H}, the Galactic Center ($A(V)\simeq 23-31$) \citep{1998A&A...337..261S,2000ApJ...537..749C, 2004ApJ...609..826K, 2006ApJ...637..774C, 2013ApJ...770...78C}, or Wolf–Rayet stars ($6 \lesssim A(V)\lesssim 13$) \citep{1998A&A...337..261S, 2001ApJ...550L.207C}, most of which likely contain dense as well as diffuse clouds. \cite{1991ApJ...371..607S} and \cite{1994ApJ...437..683P} studied the 3.4\,\micron\ hydrocarbon feature in a total of 15 sightlines with $3.9 \leq A(V) \leq 37$. More recently, \cite{2021ApJ...916...33G} studied the 10 and 20\,\micron\ silicate features in a sample of 16 sightlines with $1.8 \leq A(V) \leq 4.6$, and \cite{2024ApJ...963..120S} studied the silicate features for a sample of 49 sightlines with $ 5 \leq A(V) \leq 30$. However, none of these studies compared the extinction feature properties with elemental abundances in the dust. For sightlines with high $A(V)$s ($\gtrsim3$), it is not feasible to obtain elemental abundances from the gas absorption lines in UV spectra, because the high UV extinction precludes obtaining a high enough signal-to-noise ratio (SNR) and sufficient spectral resolution in the UV spectra, even with HST. That is why, to date, most available elemental abundance and depletion measurements are in sightlines with relatively low dust column densities ($A(V)<2.8$) \citep[e.g.,][]{2009ApJ...700.1299J}. However, the low amounts of dust in these sightlines result in extremely weak (or even undetectable) IR extinction features. A systematic study combining elemental abundances, and IR dust extinction feature properties for the same sample of sightlines has thus so far not been feasible. With MEAD, we are able to undertake this endeavor for the first time, thanks to the unprecedented sensitivity of JWST, for nine truly diffuse sightlines with $1.2 \leq A(V) \leq 2.5$. We note that some comparisons have been explored between UV extinction curves and elemental abundances by e.g., \cite{2010A&A...517A..45V}, \cite{2012ApJ...760...36P}, \cite{2016AJ....151..143H} and \cite{2021ApJS..257...63Z}.

The WISCI (Webb Investigation of Silicates, Carbons and Ices) project (Zeegers et al., in prep.) is studying dust extinction in a sample of 12 (diffuse) Milky Way sightlines.
The WISCI project is complementary to the MEAD project in two major ways: 1) WISCI uses sightlines with larger $A(V)$s ($\sim$ 4--8) compared to MEAD (1.2--2.5), and hence likely probes different ISM environments, and 2) WISCI focuses on the detailed profiles of the dust extinction features and the comparison with laboratory experiments, whereas MEAD combines extinction and abundance measurements for the same sightlines to constrain the dust grain properties. 

This paper is the first of a series on the MEAD project. In this paper, we present the IR spectra obtained with JWST, and measure the extinction features that are present in those spectra. In Sec.~\ref{sec:data}, we explain what sample of sightlines was used for this work, as well as the JWST data reduction. Sec.~\ref{sec:features} outlines the measurement and fitting of the extinction features. We show the results for the 10\,\micron\ silicate feature and compare them with other sightline properties in Sec.~\ref{sec:results}, and conclude in Sec.~\ref{sec:conclusions}.

\section{MEAD Sample and Data\label{sec:data} collection}

\subsection{Sample of sightlines}

As the goal of MEAD is to combine the information obtained from multi-wavelength (UV+IR) extinction, elemental abundance measurements in the dust, and hydrogen column densities, we require a sample for which all of this is available. To select our sample, we started from the 75 Milky Way sightlines with available UV extinction curves from \cite{2009ApJ...705.1320G}, and \ion{H}{1} and H\textsubscript{2} measurements that were later published by \cite{2023ApJ...944...33V}. As mentioned in the Introduction, elemental abundances in dust are derived from the elemental column densities measured from gas absorption lines in UV spectra. We thus had to limit our sample to those sightlines with a sufficiently large hydrogen column density (log($N(\text{H})) \gtrsim 21.5$) so that the gas absorption lines were strong enough to detect and measure. With the aim of covering as wide a parameter space as possible, we selected sightlines with a broad range in $R(V)$ (2.6--5), tracing the average grain size, and a range in molecular hydrogen fraction  $f(\text{H}_2)\ (=2N(\text{H}_2)/N(\text{H})$) (0.1--0.7), tracing the environmental conditions. To enable a statistical study of the relation between extinction, elemental abundances and environment, we selected a total of 19 sightlines from the original 75 sightlines, thereby optimizing the use of archival HST data, and limiting the request for new observations to measure elemental abundances.

In order to measure NIR/MIR extinction curves from observed spectra, it is necessary to approximate the intrinsic stellar spectrum with stellar atmosphere models. For stars with strong stellar winds (with steeply rising spectra and numerous emission lines at IR wavelengths) this is very challenging. We therefore only observed the nine (out of 19) sightlines with JWST that did not show any evidence of strong stellar winds in their Spitzer IRS (InfraRed Spectrograph) spectra (when available). For stars without IRS spectra, we used the 2MASS K\textsubscript{S} band, and WISE 22\,\micron\ or MIPS 24\,\micron\ photometry to calculate their K\textsubscript{S}--W and/or K\textsubscript{S}--M color, which is a good diagnostic for stellar winds \citep{2021ApJ...916...33G}. Only stars with K\textsubscript{S}--W or K\textsubscript{S}--M \textless 0.6 were included, resulting in the sample of nine stars that we observed with JWST (GO PID 2459; PI: Decleir), and that are studied in this paper.

The sample of sightlines used in this work is listed in Table~\ref{tab:stars}, with their properties.

\begin{deluxetable}{l|cCCC}
\tablecaption{MEAD sightlines used in this work. \label{tab:stars}}
\tablehead{\colhead{star} & \colhead{spectral type} & \colhead{$A(V)$} & \colhead{$R(V)$} & \colhead{$f(\text{H}_2)$}}
\startdata
HD014434 & O5.5V & 1.22 & 2.89 & 0.23 \\
HD038087 & B3II & 1.33 & 4.97 & 0.23 \\
HD073882 & O8.5IV & 2.46 & 3.58 & 0.67 \\
HD147888 & B3V & 1.97 & 4.08 & 0.11 \\
HD152249 & OC9Iab & 1.58 & 3.38 & 0.15 \\
HD203938 & B0.5IV & 2.35 & 3.18 & 0.40 \\
HD206267 & O6V+O9V & 1.26 & 2.66 & 0.44 \\
HD207198 & O8.5II & 1.54 & 2.68 & 0.42 \\
HD216898 & O9V & 2.50 & 2.98 & 0.31
\enddata
\end{deluxetable}

\subsection{NIRCam grism\label{sec:nircamdata}}
As the sources in our sample exceed the JWST NIRSpec (Near Infrared Spectrograph) brightness limits, the only option for obtaining spectra between 2.5 and 4~\micron\ was to utilize the JWST NIRCam (Near Infrared Camera) grism time series (TSO) mode \citep{2017JATIS...3c5001G}. 
Data were taken with the F322W2 filter, resulting in spectral coverage from $\sim$2.4--4~\micron\ at a resolution $\lambda / \Delta\lambda \sim 1400$.

The initial (``stage 1", ramps-to-slopes) data reduction was performed using custom software developed during instrument design and testing.
Processing steps included interpixel capacitance and reference correction,
dark correction, adaptive saturation detection, linearity correction, cosmic ray and snowball/cluster detection, ramp fitting, and flat fielding.

In the TSO template, we are restricted to a fixed field point (i.e., the grism spectra will always appear at a fixed location on the detector)
and dithers are not supported. These restrictions, along with field crowding, required some custom steps in the data analysis to optimize the spectral extraction:
\begin{itemize}
    \item Use of an empirical point spread function (PSF) for spectral extraction
    \item Use of a custom sensitivity curve for flux calibration
\end{itemize}

As some of our sources have nearby companions that ``contaminate" the object spectrum, we constructed a 
two dimensional empirical PSF for extractions, using isolated sources in our program and the WISCI program,
along with calibration data obtained at the TSO field point during commissioning (PID COM 1076; PI: Prizkal). 
The 2D empirical PSF is normalized in the cross-dispersion direction (columns in detector space); over the range covered in the cross-dispersion direction, the wavelength is roughly constant and we assign a single wavelength corresponding to the center of the spectral trace to all pixels in that detector column (see below). 
To extract the program object spectrum, at each detector column (constant wavelength in this approximation), the empirical PSF was fit to the program object and the flux for the object was taken as the integral of the scaled empirical PSF over the extraction box. 

For flux calibration we elected to utilize an empirical calibration for two reasons - firstly, systematic detector artifacts introduced by the fixed pointing in detector space and secondly, apparent differences in the response curve at the TSO field point when compared to the standard grism response curve. We restricted the P330-E calibrator data to only those data obtained directly at the TSO field point during commissioning (PID 1076). Those calibrator data were extracted in exactly the same way as the program objects and converted to a response curve (DN/s/MJy) using the latest CALSPEC\footnote{\url{https://www.stsci.edu/hst/instrumentation/reference-data-for-calibration-and-tools/astronomical-catalogs/calspec}} model for P330-E \citep{2017AJ....153..234B,2020AJ....160...21B,2022AJ....164...10B}. Computed in this fashion, the response curve can be directly applied to the identically extracted program spectra in pixel space without smoothing, mitigating both concerns. 

For the wavelength calibration, we used the wide field slitless spectroscopy (WFSS) calibration defined in the \texttt{nircam\_grism} package\footnote{\url{https://github.com/fengwusun/nircam_grism}} \citep{2023ApJ...953...53S}. The calibration maps an image field point (e.g., the TSO field point) to a solution for wavelength as a function of x and y detector coordinates along the location of the source trace. We verified the wavelength calibration by extracting the JWST wavelength calibrator IRAS\,05248-7007 observed at the TSO field point (PID 1076) using an identical procedure as for our program objects and applying the wavelength solution. Line positions were reproduced to 0.00001~\micron\ across the band. Additionally, for MEAD stars with measurable lines, we verified the expected line positions.

\subsection{MIRI MRS\label{sec:miridata}}
We observed our nine sightlines with the JWST MIRI (Mid-Infrared Instrument) Medium Resolution Spectrometer (MRS) \citep{2015PASP..127..646W,2023A&A...675A.111A} in all four channels and all three grating settings, covering wavelengths between 5 and 28\,\micron.

The MIRI MRS observations were reduced using the JWST pipeline (version 1.12.5) with reference files from the calibration data system pmap 1150. The standard reduction was done with some small adjustments. The residual fringe correction step was skipped in the second stage (CALWEBB\_SPEC2) as these fringes are corrected later in the pipeline in the extracted 1D spectra. In the third stage (CALWEBB\_SPEC3), the outlier detection in the cube building step was enabled. In addition, the auto-centering and residual fringe correction options were used in the spectral extraction step.
Finally, the MRS spectral leak at $\sim$12.3~\micron\ \citep{2023A&A...673A.102G} was subtracted from the extracted spectrum based on the observed spectrum at $\sim$6.1~\micron, scaled appropriately for the known leak profile.

The twelve MIRI MRS segments were merged into a single continuous spectrum using an order-dependent multiplicative correction factor measured using the overlap between orders.
Noise spikes (likely due to residual cosmic ray hits) were removed using a simple sigma-clipping algorithm, and the resulting cleaned spectra were visually inspected to make sure real emission and absorption lines were not affected.

\begin{figure*}[ht]
\centering
\includegraphics[width=0.8\textwidth]{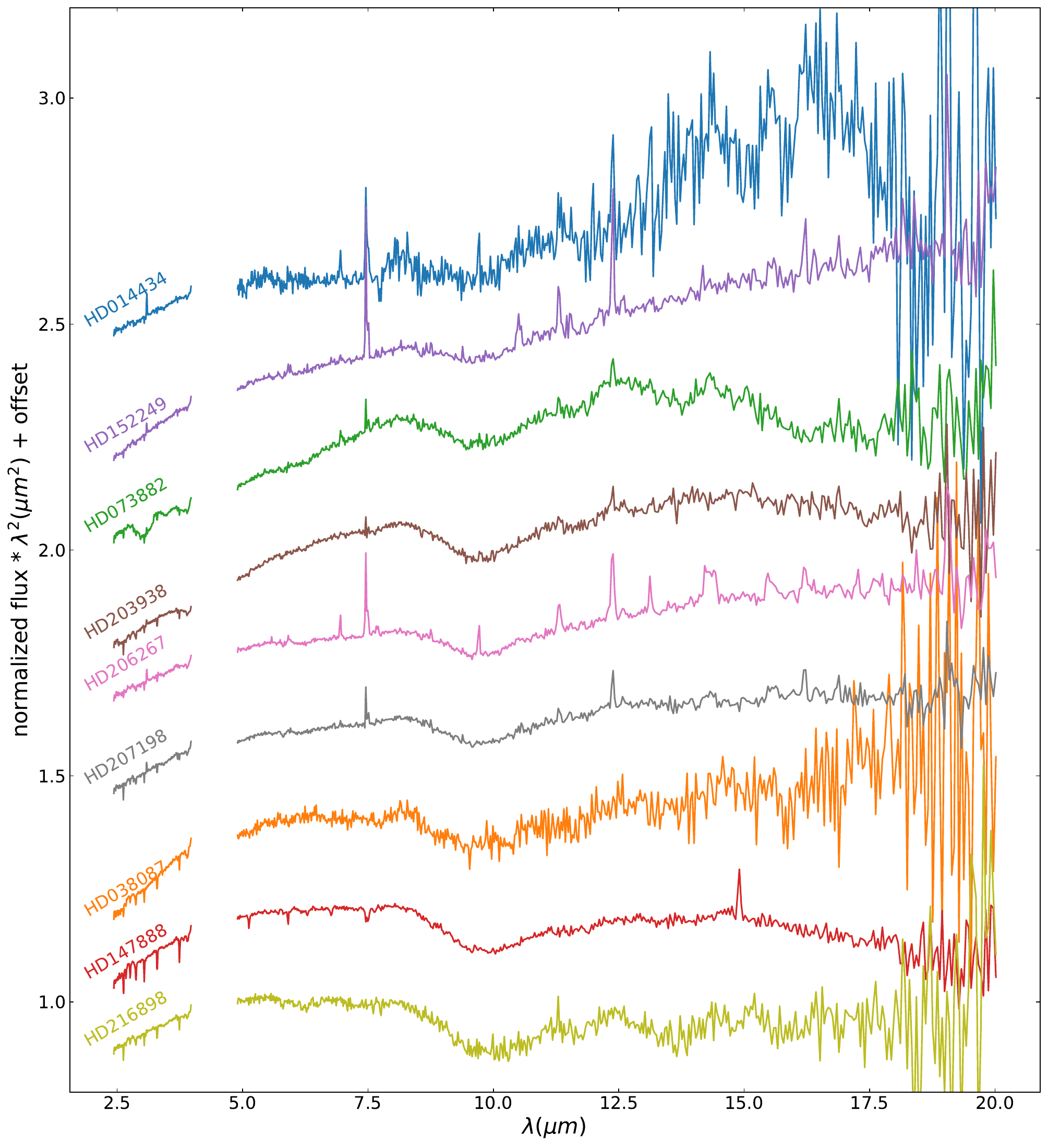}
\caption{NIRCam and MIRI spectra (rebinned to a resolution $\lambda/\Delta\lambda = 400$) for all nine stars in MEAD that were observed with JWST, multiplied by $\lambda^2$ to flatten out the decreasing Rayleigh-Jeans tail of the stellar spectrum at these wavelengths, normalized to the mean flux*$\lambda^2$ between 5 and 7.5\,\micron, and ordered from flattest (bottom) to steepest (top) spectrum for convenient visualization. 
 \label{fig:spectra}}
\end{figure*}

The JWST spectra presented in this paper were obtained from the Mikulski Archive for Space Telescopes (MAST) at the Space Telescope Science Institute. The used observations can be accessed via \dataset[10.17909/dp6s-rd16]{http://dx.doi.org/10.17909/dp6s-rd16}. All reduced and calibrated NIRCam and MIRI MEAD spectra are shown in Fig.~\ref{fig:spectra}, and electronically available\footnote{\url{https://doi.org/10.5281/zenodo.14286122}} \citep{decleir_2024_14286122}. For the purpose of this paper, we re-binned the spectra to a resolution $\lambda / \Delta\lambda = 400$. 

\section{Measuring IR extinction features\label{sec:features}}

\subsection{Silicate feature}\label{sec:fit_sil}
By far the strongest feature in the JWST spectra is the 10\,\micron\ silicate feature. We fit the continuum flux around the feature (using the fluxes at 7.9--8.1 and at 12.6--12.8\,\micron) with a line using the \texttt{Linear1D} Astropy \citep{astropy:2013,astropy:2018,astropy:2022} model. We first multiplied the spectra (in Jy) by $\lambda^2$ to remove the strong Rayleigh-Jeans decrease expected for stars in this wavelength range. This mostly flattens the spectrum around the feature, which justifies fitting the local continuum with a line.\footnote{Not multiplying the spectrum by $\lambda^2$ results in a significant overestimate of the strength of the feature (with peak optical depths $\sim3$ times larger) when using a line to fit the local continuum due to the intrinsic curvature of the stellar spectrum.} Prominent stellar lines were masked from the spectra before the fitting. We used the Astropy \texttt{FittingWithOutlierRemoval} fitter in combination with the \texttt{LinearLSQFitter} to remove outliers ($>3\sigma$) from the fitting. We then normalized the spectrum ($F(\lambda)_{\text{norm}}$) and calculated the optical depth $\tau$ at every wavelength $\lambda$ (between 7.9 and 12.8\,\micron) as follows:
\begin{equation}
\tau (\lambda) = \text{ln} (1 / F(\lambda)_{\text{norm}}) \label{eq:tau}
\end{equation}

\begin{figure}[ht]
\centering
\includegraphics[width=\columnwidth]{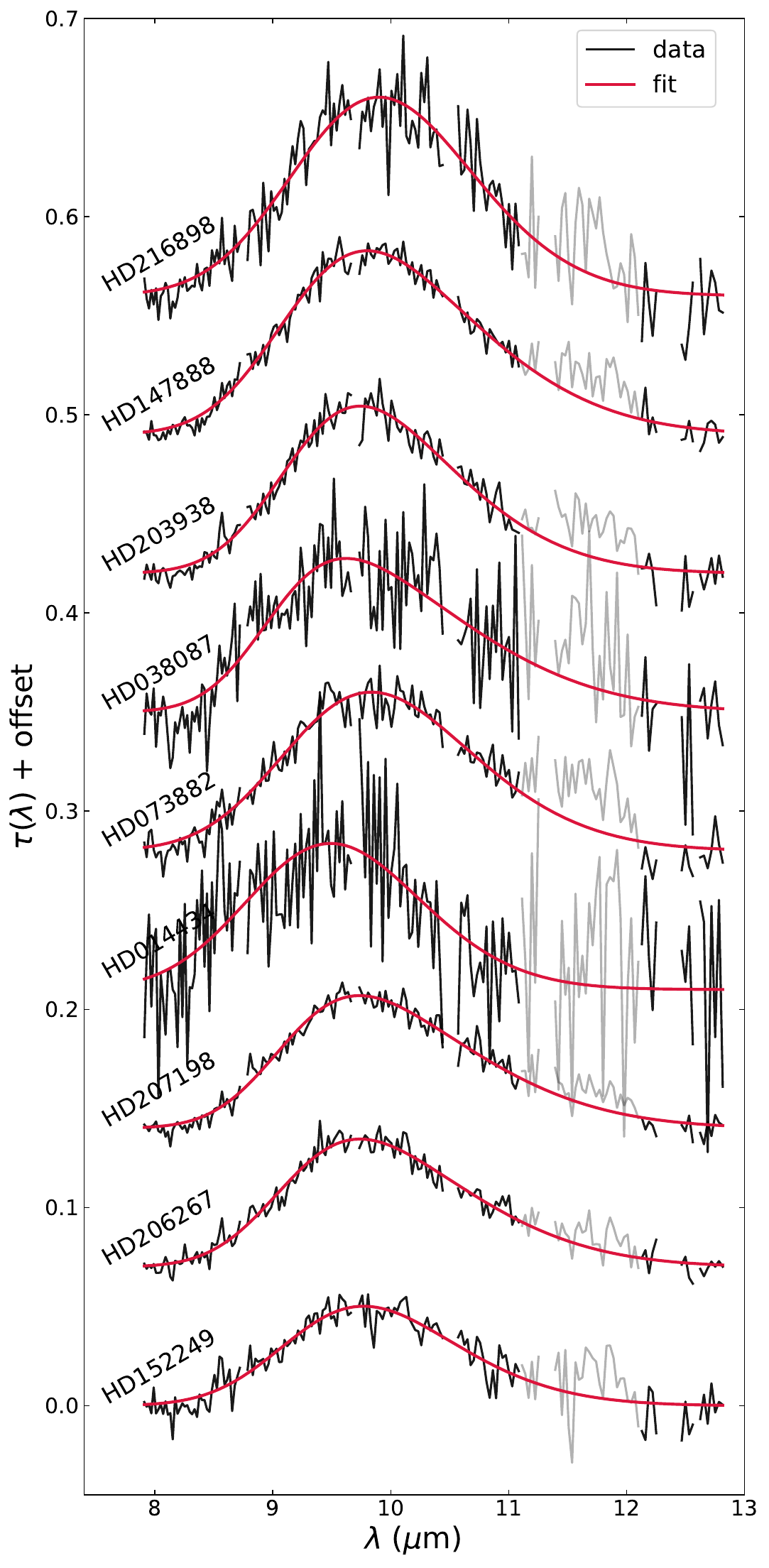}
\caption{Silicate features, ordered from weakest (bottom) to strongest (top). The skewed Gaussian fits are shown in red.
 \label{fig:si_features}}
\end{figure}

The optical depths for all sightlines are shown in Fig.~\ref{fig:si_features}. We fit the feature with a skewed Gaussian profile because the feature is not symmetric. We used the SciPy \citep{2020SciPy-NMeth} \texttt{skewnorm}\footnote{\url{https://docs.scipy.org/doc/scipy/reference/generated/scipy.stats.skewnorm.html}} probability density function (PDF), which is defined as:
\begin{equation}
f(x) = 2\phi (x)\Phi (\alpha x) \label{eq:pdf}
\end{equation}
with a shape parameter $\alpha$, and

\begin{equation}
\phi (x)={\frac {1}{\sqrt {2\pi }}}e^{-{\frac {x^{2}}{2}}}
\end{equation}
the standard Gaussian PDF, and

\begin{equation}
\Phi (x) = \int _{-\infty }^{x}\phi (t)\ \mathrm {d} t = {\frac {1}{2}}\left[1+ \text{erf} \left({\frac {x}{\sqrt {2}}}\right)\right]
\end{equation}
the cumulative distribution function, where ``erf" is the error function. When replacing $x$ by $\frac {x-\xi }{\omega}$, we add a location $\xi$ and a scale $\omega$ to shift the peak and scale the width of the profile. This PDF is normalized (by definition), so that the integrated area under the profile is equal to one. In order to fit the observed silicate feature, we multiplied this PDF with an amplitude parameter $B$. The model that is fit to the feature is thus:
\begin{equation}\label{eq:skew}
\tau(\lambda) = B \frac{1}{\sqrt{2\pi}\omega} e^{-\frac{(\lambda-\xi)^{2}}{2\omega^2}} \left[1 + \text{erf} \left(\alpha \frac{\lambda-\xi}{\sqrt{2}\omega} \right)\right]
\end{equation}

The fitting of the feature was done in two steps. First, the Levenberg–Marquardt algorithm was used to obtain preliminary fit results for the four parameters ($B$, $\xi$, $\omega$ and $\alpha$), using the Astropy \texttt{LevMarLSQFitter}. These results were then used as initial guesses in the Markov chain Monte Carlo (MCMC) fitting with the \texttt{Emcee} python tool \citep{2013PASP..125..306F}. We used 8 walkers, each with 9000 steps after a burn in of 1000 steps to sample the parameter space.
We are taking the 50\textsuperscript{th} percentile of the posterior distribution function of every parameter as the final fit result. The fitted curves are shown as red lines in Fig.~\ref{fig:si_features}.  The asymmetric uncertainties on the fitted parameters given by the MCMC fitting were calculated as the difference between the 84\textsuperscript{th} and 50\textsuperscript{th} percentile (upper uncertainty), and between the 50\textsuperscript{th} and 16\textsuperscript{th} percentile (lower uncertainty) of the posterior distribution function. We note that two sightlines have noisy MIRI spectra (HD014434 and HD038087), and their fits are less reliable.

It is important to note that because of the mathematical definition of the skewed Gaussian (Eq.~\ref{eq:skew}), some of the fitted parameters are not physically meaningful. We therefore also derived the mode (peak wavelength) $\lambda_0$, peak optical depth ($\tau(\lambda_0)$) and full width at half maximum (FWHM) of the fitted profile. There is no analytic expression for the mode, so we used the numerical approximation derived by \cite{azzalini2014skew}:
\begin{equation}
\lambda_0 \approx \xi + \omega \left[ \sqrt{\frac{2}{\pi}} \delta - \left(1-\frac{\pi}{4}\right) 
 \frac{\left(\sqrt{\frac{2}{\pi}}\delta \right)^{3}}{1-\frac{2}{\pi}\delta^{2}} - \frac{\mathrm{sgn}(\alpha)}{2} e^{-\frac{2\pi}{|\alpha|}} \right]
\end{equation}
with
\begin{equation}
\delta = \frac{\alpha}{\sqrt{1+\alpha^{2}}}
\end{equation}

\noindent Given that the PDF in Eq.~\ref{eq:pdf} is normalized, the amplitude parameter $B$ is equal to the integrated area under the fitted profile. Finally, we will consider the shape parameter $\alpha$ as a measure of the asymmetry. A positive value for $\alpha$ corresponds to a profile that is right skewed (i.e., longer tail to the right of the peak), while a negative value corresponds to a left skewed profile (i.e., longer tail to the left of the peak), and the profile becomes a regular Gaussian for $\alpha=0$. The obtained feature properties and their uncertainties are given in Table~\ref{tab:fit_results}, together with the median values for our sample.

\begin{deluxetable*}{l|CCCCC}
\tablecaption{Derived properties of the 10\,\micron\ silicate feature, based on the MCMC fitting results. Median values for this sample are listed in the bottom row. \label{tab:fit_results}}
\tablehead{\colhead{star} & \colhead{$\lambda_0$(\micron)} & \colhead{$\tau$($\lambda_0$)} & \colhead{FWHM(\micron)} & \colhead{area(\micron) (=$B$)} & \colhead{$\alpha$}}
\startdata
HD014434 & 9.47_{-0.11}^{+0.09} & 0.071_{-0.007}^{+0.007} & 1.76_{-0.20}^{+0.24} & 0.138_{-0.013}^{+0.015} & 1.22_{-0.82}^{+1.17} \\
HD038087 & 9.63_{-0.04}^{+0.04} & 0.078_{-0.002}^{+0.002} & 1.92_{-0.06}^{+0.06} & 0.161_{-0.004}^{+0.004} & 2.95_{-0.44}^{+0.47} \\
HD073882 & 9.84_{-0.03}^{+0.03} & 0.081_{-0.001}^{+0.001} & 1.90_{-0.04}^{+0.04} & 0.164_{-0.003}^{+0.003} & 1.69_{-0.32}^{+0.29} \\
HD147888 & 9.80_{-0.01}^{+0.01} & 0.091_{-0.001}^{+0.001} & 1.94_{-0.02}^{+0.02} & 0.194_{-0.002}^{+0.002} & 2.20_{-0.14}^{+0.15} \\
HD152249 & 9.77_{-0.03}^{+0.03} & 0.051_{-0.001}^{+0.001} & 1.74_{-0.04}^{+0.04} & 0.094_{-0.002}^{+0.002} & 1.76_{-0.31}^{+0.31} \\
HD203938 & 9.74_{-0.02}^{+0.01} & 0.084_{-0.001}^{+0.001} & 1.73_{-0.02}^{+0.02} & 0.157_{-0.002}^{+0.002} & 2.10_{-0.16}^{+0.17} \\
HD206267 & 9.74_{-0.01}^{+0.01} & 0.064_{-0.001}^{+0.001} & 1.85_{-0.02}^{+0.02} & 0.129_{-0.001}^{+0.001} & 2.43_{-0.11}^{+0.11} \\
HD207198 & 9.74_{-0.02}^{+0.01} & 0.068_{-0.001}^{+0.001} & 1.88_{-0.03}^{+0.03} & 0.136_{-0.002}^{+0.002} & 2.64_{-0.16}^{+0.16} \\
HD216898 & 9.90_{-0.03}^{+0.03} & 0.101_{-0.002}^{+0.002} & 1.83_{-0.04}^{+0.04} & 0.196_{-0.004}^{+0.004} & 1.28_{-0.52}^{+0.36} \\
\hline
median & 9.74 & 0.078 & 1.85 & 0.157 & 2.10
\enddata
\end{deluxetable*}

It has to be noted that we also tried to fit the feature with a modified Drude profile, as was done by \cite{2021ApJ...916...33G}. We found that this profile did not fit the observations well. \cite{2021ApJ...916...33G} used a combination of a power law for the continuum and modified Drude profiles for the two silicate features at 10 and 20\,\micron\ to fit their measured extinction curves. As can be seen in their figure 9, the extinction beyond the 10\,\micron\ feature does not return to the continuum level, as the 20\,\micron\ feature already starts affecting the extinction. As the longer wavelength region ($\gtrsim 18$\,\micron) in our MIRI spectra is quite noisy, we cannot constrain the 20\,\micron\ feature directly from the spectra. We thus measured the 10\,\micron\ feature in isolation and as explained before, we fit a line to the local ``continuum" level. This forces the optical depth to go back to zero around 13\,\micron\ (see Fig.~\ref{fig:si_features}). We believe this could explain why a modified Drude profile, which has a much longer tail than a (skewed) Gaussian profile, does not fit our data well. In future work (MEAD paper III, Decleir et al., in prep.), we will use stellar atmosphere models to measure the complete NIR--MIR extinction curve. This will give us a more accurate constraint on the underlying continuum extinction. We will then re-evaluate the suitability of a modified Drude to fit the silicate feature.

In all sightlines, an additional feature can be observed on top of the main 10\,\micron\ feature between $\sim$11.1 and $\sim$12.1\,\micron\ (shown in gray in Fig.~\ref{fig:si_features}). The origin of this feature is unclear, and possible carriers are discussed in the next section. A detailed study of the strength, shape and origin of this extra feature is beyond the scope of this work, and the wavelength region between 11.1 and 12.1\,\micron\ has been masked from the fitting of the main silicate feature.

\subsection{Unidentified feature between 11 and 12\,\micron}
\cite{2021ApJ...916...33G} did not observe an additional feature between 11 and 12\,\micron\ in the Spitzer data of their sightlines. This could indicate that the feature we observe here is caused by a JWST/MIRI instrumental artifact. However, the strength of this extra feature seems to (more or less) scale with the strength of the main silicate feature, suggesting a dusty nature.

Crystalline forsterite (Mg\textsubscript{2}SiO\textsubscript{4}) causes an absorption band that peaks between 11.4 and 11.5\,\micron\, depending on its exact properties as prepared in the laboratory \citep[e.g.,][]{2001A&A...378..228F,2004A&A...413..395B}.
However, those materials also cause peaks around other wavelengths (such as 9.8, 10.3, 10.5 and 10.7\,\micron), which we do not observe, although the strengths of these features depend on the grain shape, and might be too weak to detect in our spectra. \cite{2016MNRAS.457.1593W} and \cite{2020MNRAS.493.4463D} detected an absorption feature at 11.1\,\micron\ in the spectra of evolved stars, embedded Young Stellar Objects (YSOs), paths toward the Galactic Center, and in the diffuse ISM, and also attributed it to crystalline forsterite. However, the feature we observe in our spectra seems to peak at longer wavelengths.

Another possible carrier of this feature could be Si\sbond C stretching of silicon carbide (SiC), which causes a feature around 11.3\,\micron, that has been detected in emission and in absorption toward carbon-rich evolved stars \citep[e.g.,][]{1997MNRAS.288..431S}. \cite{2022MNRAS.509.5231C} argue that SiC dust could be a significant constituent of ISM dust since it is generally believed that carbon stars inject a considerable amount of dust into the ISM. However, according to \cite{2022MNRAS.509.5231C}, the 11.3\,\micron\ absorption feature of SiC has never been seen in the ISM before \citep[see also][]{1990MNRAS.244..427W}. If the feature we observe in our spectra is indeed related to SiC, this could potentially solve a long-standing mystery around the lack of SiC in the ISM.

Also polycyclic aromatic hydrocarbons (PAHs) could be responsible for a feature around these wavelengths. \cite{2000ApJ...544L..75B} observed an absorption band at 11.25\,\micron\ in the embedded YSO Monoceros R2 IRS 3, and identified it with a C\sbond H out-of-plane vibrational mode of PAH molecules. However, this PAH feature would likely be accompanied by other bands (such as the C\sbond C mode at 7.7\,\micron), which we do not observe (although this feature could be too weak to be detectable in our spectra).

Other possible candidates for a feature between 11 and 12\,\micron\ that have been suggested in the literature include: carbonates (peak near 11.4\,\micron) \citep[e.g.,][and references therein]{2016MNRAS.457.1593W}, polycrystalline graphite (peak near 11.5\,\micron) \citep{1984ApJ...277L..71D,2016ApJ...831..109D, 2020ApJ...895...38H}, water ice (peak near 11.6\,\micron) \citep[e.g.,][]{2024A&A...681A...9R}, and nano-sized silicate grains \citep[e.g.,][]{2023FaDi..245..609Z}.

In conclusion, there are at least seven plausible candidates for the feature observed between 11 and 12\,\micron, if it is not caused by an instrumental artifact. This feature (and its possible carriers) will be revisited when measuring the complete NIR--MIR extinction curves in future work (MEAD paper III, Decleir et al., in prep.).

\subsection{Carbonaceous features}
\label{sec:carbon}
As mentioned in the Introduction, previous studies have observed an extinction feature caused by aliphatic hydrocarbons around 3.4\,\micron\ \citep[e.g.,][]{1991ApJ...371..607S}. This feature is generally attributed to hydrocarbons in the diffuse ISM, but has so far only been observed in sightlines with relatively high $A(V)$s, such as toward the hypergiant Cyg OB2-12 ($A(V)\simeq10$) \citep{1990MNRAS.243..400A,1998A&A...337..261S,2020ApJ...895...38H}; supergiants ($3.9\leq A(V)\leq 10$), Wolf-Rayet stars ($5.2\leq A(V)\leq 12.8$) and the Galactic Center ($A(V)\sim 31$) \citep{1994ApJ...437..683P}; highly reddened stars in the Galactic Plane ($5.8\leq A(V) \leq11.9$) \citep{1996AJ....112..235I}; highly reddened ($9<A(V)<15.8$) early-type stars \citep{2003MNRAS.341.1121R}; the YSO IRAS\,18511+0146 ($A(V)\sim7$) \citep{2012A&A...537A..27G}; the Quintuplet in the Galactic Center ($A(V)\simeq 29$) \citep{2013ApJ...770...78C}; and several other high $A(V)$ sightlines in other works.

\begin{figure*}[ht]
  \plottwo{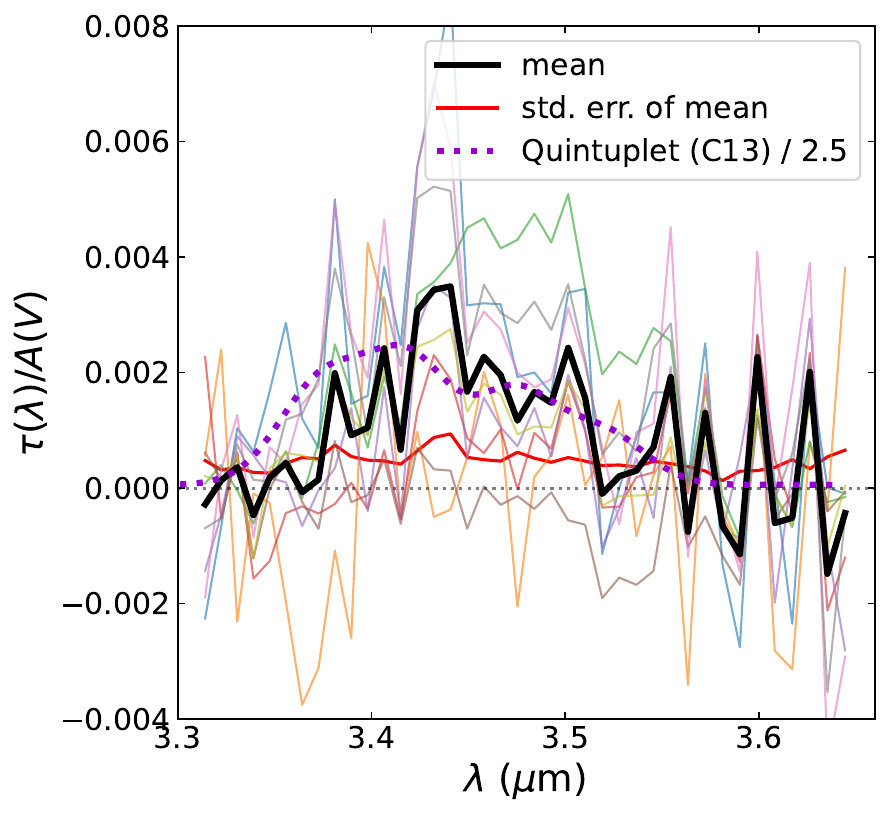}{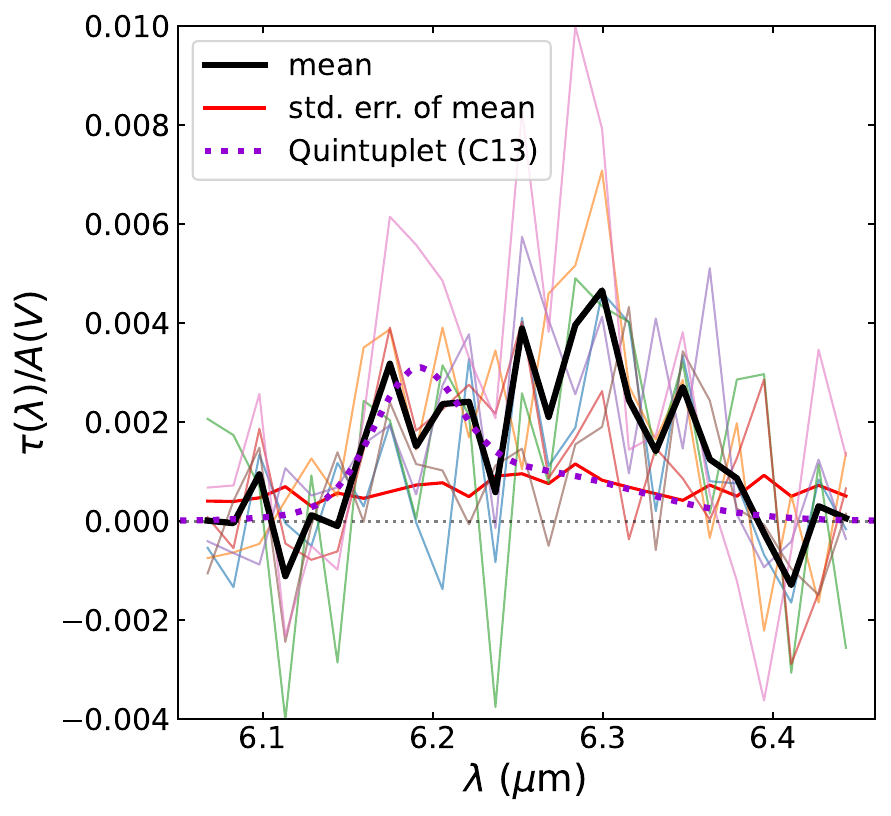}  
  \caption{3.4\,\micron\ (left) and 6.2\,\micron\ (right) hydrocarbon features, normalized to $A(V)$ for all sightlines (in color). The average is shown in black, and the standard error of the mean in red. The purple dotted line shows the feature for the Quintuplet as fit by \cite{2013ApJ...770...78C} (C13). In the left plot, we divided the C13 profile by a factor of 2.5. The two sightlines with noisy MIRI spectra are not included in the right plot.} \label{fig:tau34}
\end{figure*}

In our purely diffuse, low $A(V)$ sightlines, it is very challenging to detect this weak feature. Nevertheless, an extinction feature is clearly visible in the spectra of three of our sightlines around 3.4\,\micron\ (HD073882, HD207198 and HD216898). To improve the SNR, we averaged all observed MEAD spectra as follows. We fit the local continuum around the 3.4\,\micron\ region with a line (after multiplying the fluxes by $\lambda^2$ to flatten out the Rayleigh Jeans decrease), normalized the spectra and calculated the optical depths in the same way as for the silicate feature (see Eq.~\ref{eq:tau}). We then divided the optical depths for each sightline by its $A(V)$ in order to give each sightline an equal weight in the average. The normalized optical depths for each sightline are shown in Fig.~\ref{fig:tau34} (left), together with the average of all sightlines (in black), and the standard error of the mean (in red). The feature is clearly detected in the average. No profile was fit to the feature, because the SNR in the NIRCam spectra is not high enough to constrain this weak feature. Instead, here, we only report the maximum optical depth, found at $\lambda_0\sim3.44\,\micron$, and the standard error of the mean at that wavelength: $\tau(3.4\,\micron)/A(V) \approx 0.0035 \pm 0.0009$, which corresponds to a $\sim4\sigma$ detection. To our knowledge, this is the first detection of the 3.4\,\micron\ feature in sightlines with $A(V)\leq 2.5$. Our average $A(V)/\tau(3.4\,\micron)= 286_{-61}^{+105}$ agrees well with results reported in the literature for the local diffuse ISM: $240\pm40$ \citep{1991ApJ...371..607S}; $250\pm40$ \citep{1994ApJ...437..683P}; and 333 \citep{1990MNRAS.243..400A,1996AJ....112..235I}. Within our uncertainties, it also agrees with the more recent value $A(V)/\tau(3.4\,\micron)= 232_{-24}^{+30}$ for Cyg OB2-12 measured by \cite{2020ApJ...895...38H}.

We added the fitted profile (which is the sum of four Gaussians) from \cite{2013ApJ...770...78C} for the Quintuplet as a purple dotted line, omitting their aromatic hydrocarbon component at 3.3\,\micron. As our spectra have a strong stellar line around that wavelength, we cannot use those data to search for an aromatic component. We divided the fitted profile by a factor of 2.5 to more or less match the strength of the feature in our sample average. Overall, the shape of the average feature is in reasonable agreement with the shape observed for the Quintuplet. The discrepancy at the short-wavelength side of the feature is likely caused by different assumptions in the continuum fitting. It is also possible that the ratios between the four separate aliphatic components of this feature are different in our diffuse sightlines compared to the dense sightline toward the Quintuplet in \cite{2013ApJ...770...78C}.

We did the same analysis around the 6.2\,\micron\ region, where an olefinic and aromatic hydrocarbon feature has been detected in previous studies in high $A(V)$ sightlines, such as late-type Wolf-Rayet stars ($6.4\leq A(V)\leq12.8$), the hypergiant Cyg OB2-12 ($A(V)\simeq10$) and the Galactic Center ($A(V)\sim23-29$) \citep[e.g.,][]{1998A&A...337..261S, 2000ApJ...537..749C, 2001ApJ...550L.207C, 2013ApJ...770...78C, 2020ApJ...895...38H}.
The individual and average $A(V)$-normalized optical depths are shown in Fig.~\ref{fig:tau34} (right), excluding the two sightlines with noisy MIRI spectra (HD014434 and HD038087).
We again added the fitted profile (which is the sum of two Gaussians) from \cite{2013ApJ...770...78C} for the Quintuplet as a purple dotted line. Note that we did not scale the fitted profile to match our average feature in this case. The shape of our feature matches well around 6.2\,\micron, however, we detect an excess around 6.3\,\micron\ in our average. We note that a stellar line around 6.3\,\micron\ might cause an issue. However, it is also possible that the ratio between the aromatic (peak $\sim6.25\,\micron$) and olefinic (peak $\sim6.19\,\micron$) components of this feature is different in our diffuse sightlines compared to the dense sightline toward the Quintuplet in \cite{2013ApJ...770...78C}. If we focus on the bump around 6.2\,\micron, the maximum optical depth is found at $\lambda_0\sim6.17\,\micron$ and is $\tau(6.2\,\micron)/A(V) \approx 0.0032 \pm 0.0006$, which corresponds to a $\sim5\sigma$ detection. To our knowledge, this is the first (tentative) detection of the 6.2\,\micron\ feature in diffuse sightlines with $A(V)\leq 2.5$. Our average $A(V)/\tau(6.2\,\micron)= 314_{-49}^{+72}$. Values in the literature wildly range from $\sim177$ to $\sim453$ \citep{1998A&A...337..261S,2000ApJ...537..749C,2001ApJ...550L.207C,2013ApJ...770...78C}, but are mostly measured in dense sightlines, which makes a comparison with our diffuse sightlines more difficult. Our value is significantly smaller than the value $A(V)/\tau(6.2\,\micron)= 464_{-20}^{+22}$ for Cyg OB2-12 measured by \cite{2020ApJ...895...38H}.

We also detected an absorption feature around 5.8\,\micron, which could be caused by carbonyl (C\dbond O) groups on interstellar dust grains \citep[see e.g.,][]{1996ApJ...461..210T,2013ApJ...770...78C}. However, we are suspicious that this feature might be an instrumental artifact, because we also observed it in the spectrum of calibration star 10 Lac, and hence did not further investigate this feature.

A more detailed study of the carbonaceous features will be done in future work (MEAD paper III, Decleir et al., in prep.), where we will use stellar atmosphere models to better constrain the continuum level around these features, as well as to better account for stellar lines.

\subsection{Water ice or trapped water feature}\label{sec:ice}
Extinction features caused by water ice, for example the O\sbond H stretching feature peaking around 3\,\micron,  are generally accepted to be detectable only in sightlines that contain a dense molecular component \citep[e.g.,][]{1997ApJ...490..729W, 2015ARA&A..53..541B}. However, \cite{2022ApJ...930...15D} reported a tentative detection ($\sim3.3\sigma$ and $3.5\sigma$, respectively) of a weak 3\,\micron\ water ice feature in the diffuse sightlines HD183143 and HD229238 (with $A(V)=3.86$ and 2.99, respectively). The strength of this feature in their average diffuse extinction curve is, however, below a $3\sigma$ detection threshold of $A(\text{ice})/A(V) = 0.0021$. They note that their analysis of this feature was challenged by the significant telluric absorption in their ground-based SpeX spectra around these wavelengths.

Simulations by \cite{2007ApJ...668..294C} showed that around one monolayer of ice can be formed on carbonaceous dust grains in diffuse/translucent clouds with $A(V) \sim2-3$, similar to the $A(V)$-values of some MEAD sightlines. In addition, experiments by \cite{2011ApJ...741L...9J} demonstrated that deuterated water ice (D\textsubscript{2}O and HDO) can be formed on amorphous silicate surfaces in conditions resembling regions of the ISM with $A(V) \sim1-3$. Finally, \cite{2021NatAs...5...78P} suggested solid-state water molecules can be trapped (strongly bound) on the surface of silicate grains. They provided evidence for the presence of trapped solid-state water in the diffuse ISM (toward Cyg OB2-12). \cite{2024ApJ...965...48P} further conclude that the detection of solid-state water in the diffuse ISM speaks for its efficient, continuous formation (most likely through reactions between hydrogen and oxygen atoms/molecules on the dust surface) which counteracts the desorption and destruction of water molecules by photons.

\begin{figure*}[ht]
  \plottwo{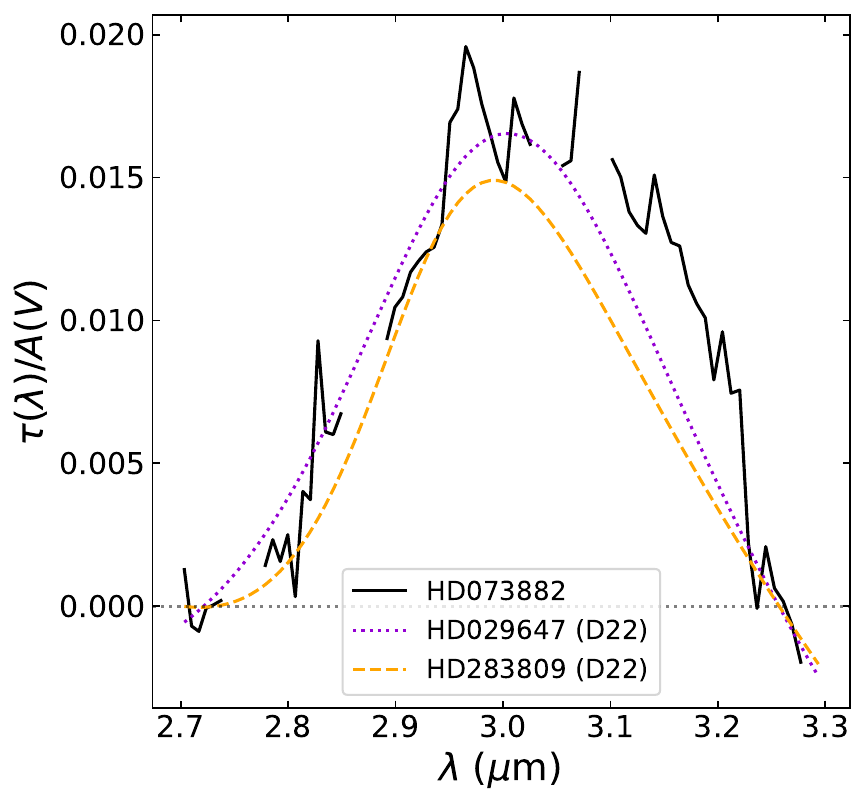}{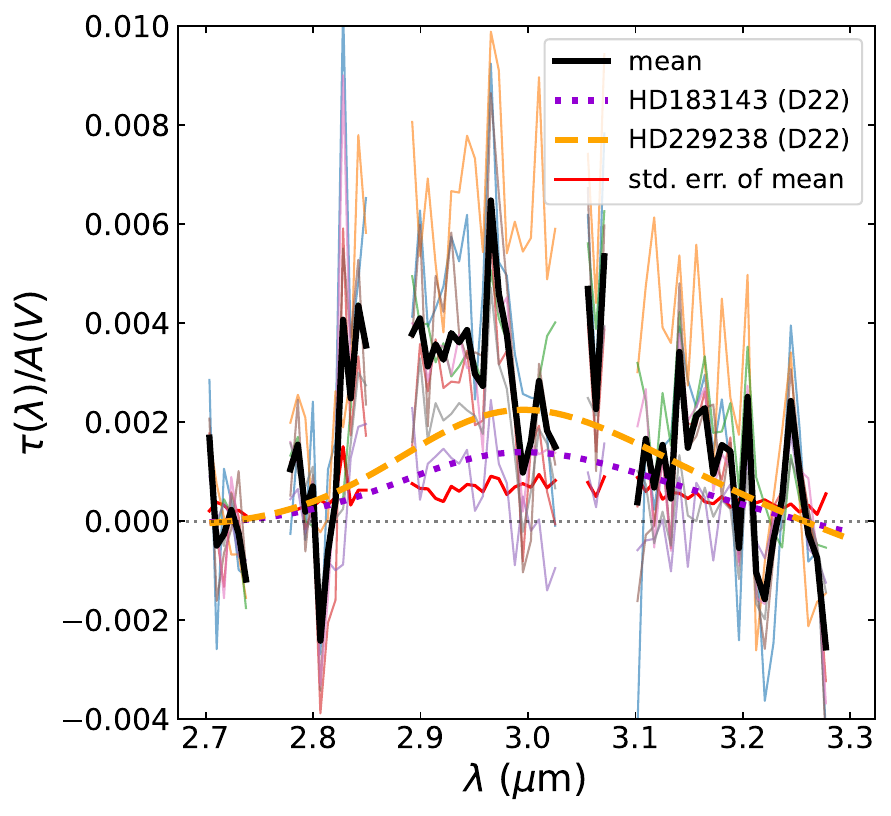}
  \caption{3\,\micron\ feature  normalized to $A(V)$ for HD073882 (left), and for the rest of the sample (right) (in color). The average is shown in black, and the standard error of the mean in red. In dashed and dotted lines, we show the re-normalized profiles from \cite{2022ApJ...930...15D} for the two dense sightlines (left), and two diffuse sightlines with a tentative ice detection (right) in their sample.} \label{fig:tau30}
\end{figure*}

In our sample, a clear 3\,\micron\ feature can be observed in HD073882. After fitting the local continuum, and calculating the optical depths as was done for the other features, we find a maximum $\tau(3\,\micron)/A(V) \sim 0.020 \pm0.001$ at $\lambda_0\sim2.97$, which corresponds to a $20\sigma$ detection (see Fig.~\ref{fig:tau30}, left). This sightline has an $A(V)=2.46$ and an $f(\text{H}_2)=0.67$. Given the relatively large fraction of $\text{H}_2$ in this sightline (the largest in our sample), the presence of water ice is not entirely unexpected. Nevertheless, to our knowledge, this is the first detection of water ice in a sightline with $A(V)\lesssim2.5$. We added the fitted profiles for the two dense sightlines (HD029647 and HD283809) from \cite{2022ApJ...930...15D}, continuum-subtracted using the same wavelengths to fit the continuum as we used for HD073882.\footnote{\cite{2022ApJ...930...15D} fit the extinction curves of these sightlines using a combination of a powerlaw for the continuum and a modified Drude profile for the 3\,\micron\ feature.} The shape of our feature is in reasonable agreement with that of HD029647, although we find a small excess on the long-wavelength side of the feature.

Averaging all other MEAD spectra in this wavelength region (i.e., excluding HD073882) in the same way as was done for the 3.4 and 6.2\,\micron\ regions (see Sec.~\ref{sec:carbon}), results in a maximum $\tau(3\,\micron)/A(V) \approx 0.0065 \pm 0.0009$ at $\lambda_0\sim2.97$, corresponding to a $\sim7\sigma$ detection (see Fig.~\ref{fig:tau30}, right). This is almost three times stronger than the feature found by \cite{2021NatAs...5...78P} in the spectrum of Cyg OB2-12. We added the fitted profiles for the diffuse sightlines with a tentative ice detection (HD183143 and HD229238) from \cite{2022ApJ...930...15D}, continuum-subtracted using the same wavelengths to fit the continuum as we used for the MEAD spectra. The average 3\,\micron\ feature seems to be stronger than in those diffuse sightlines, and its peak seems shifted towards shorter wavelengths. However, we want to caution that we masked several strong stellar lines at these wavelengths in our spectra, which can potentially influence the peak and shape of this feature. In addition, we note that there could be an instrumental artifact around these wavelengths causing the feature to look stronger than it actually is. However, no 3\,\micron\ feature was detected in the calibration star P330-E.

A more detailed study of this feature is beyond the scope of this work, as the SNR of the current data is insufficient to draw strong conclusions about its carrier. We will revisit the possible detection of water ice (or trapped solid-state water) in our sample in future work (MEAD paper III, Decleir et al., in prep.), using stellar atmosphere models to more carefully account for the stellar lines in the observed spectra.

The code that we developed to measure the extinction features, and analyze and plot the results in this work, is available as part of the \texttt{mead} package on GitHub\footnote{\url{https://github.com/mdecleir/mead/releases/tag/v1.0.0}} \citep{marjorie_decleir_2024_14291651}.

\section{Results and discussion}\label{sec:results}

\subsection{Silicate feature variations}

\begin{figure*}[ht]
\centering
\includegraphics[width=0.65\textwidth]{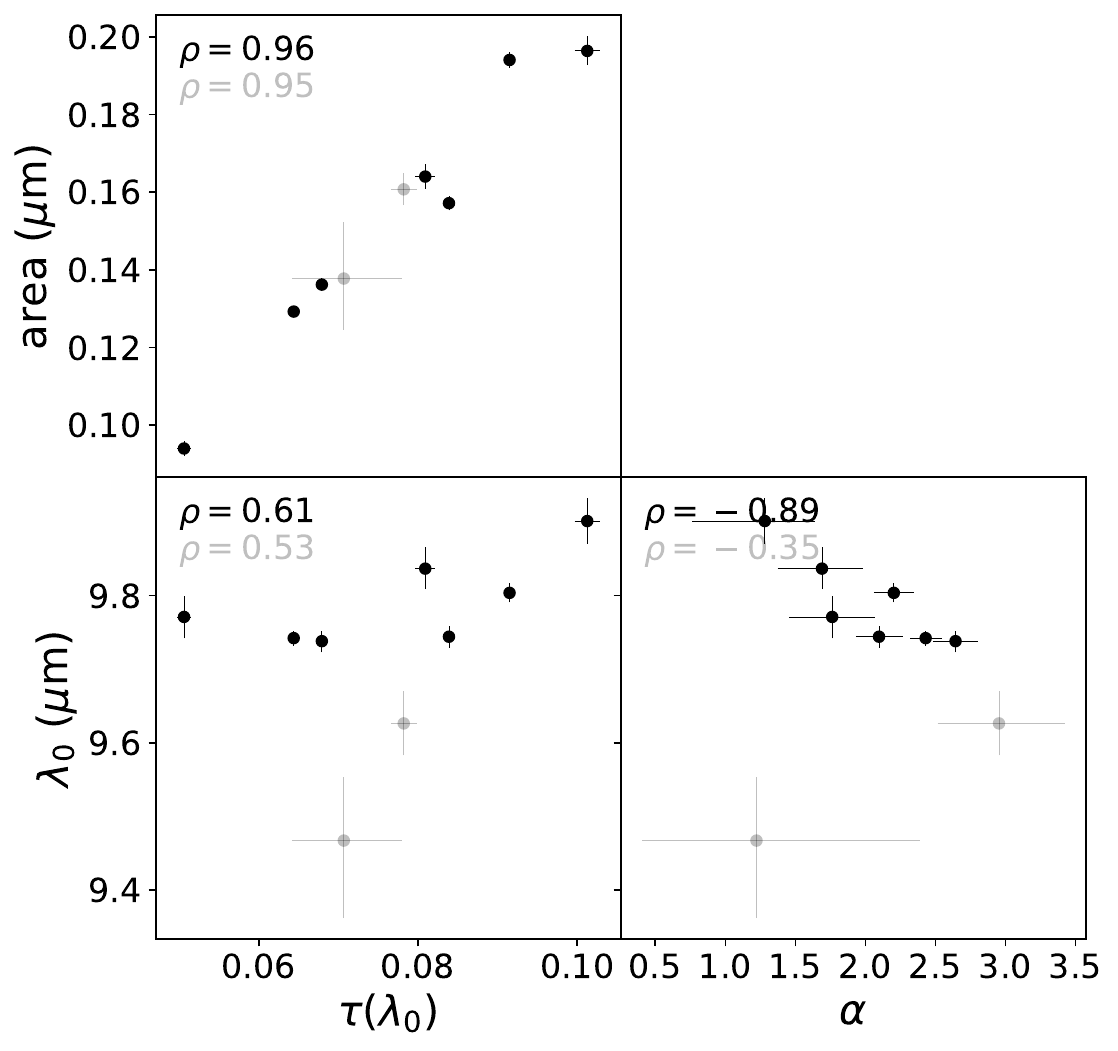}
\caption{Observed correlations between silicate feature properties. The gray data points correspond to the two sightlines with a noisy MIRI spectrum. The Spearman's rank correlation coefficients are shown in the corner of each panel (in black for only the black data points, and in gray for all the data points).
 \label{fig:sil_params}}
\end{figure*}

As explained in Sec.~\ref{sec:fit_sil}, from the fitting results we derived the properties of the silicate feature: the peak wavelength, $\lambda_0$, the peak optical depth, $\tau(\lambda_0)$, the FWHM, the asymmetry, $\alpha$, and the integrated area of the fitted profile.
As can be seen in Table~\ref{tab:fit_results}, there is a significant variation in some of these properties between the different sightlines of the MEAD sample. This indicates that different sightlines contain silicate grains with different physical and chemical properties (such as composition, stoichiometry, size, shape or crystallinity) \citep[e.g.,][]{1979ApJ...234..158D,1986Ap&SS.128...47D,1995A&A...300..503D,2011ApJ...740...93S}.
Fig.~\ref{fig:sil_params} shows correlations between some of the silicate feature parameters. In this plot (and all the following ones), the two sightlines with noisy MIRI spectra and less reliable fits (HD014434 and HD038087) are indicated in gray. The Spearman's rank correlation coefficient ($\rho$) is indicated in every panel in black for the data without the two noisy sightlines, and in gray for all data points, in this plot (and all the following ones).

There is a very strong correlation ($\rho=0.96$) between the integrated area and the peak optical depth of the feature, which is to be expected, given that the FWHM of the feature does not change much between different sightlines (see below). In other words, the total extinction in the silicate feature is well approximated by the peak optical depth. Other works \citep[e.g.,][]{2021ApJ...916...33G, 2024ApJ...963..120S}, usually only use the peak optical depth in their analysis, which seems to be well justified. Given the strong correlation between both parameters, we will also focus on the peak optical depth in the analyses and plots in the rest of this paper, in order to facilitate comparisons with literature results.

\begin{figure}[ht]
\centering
\includegraphics[width=0.9\columnwidth]{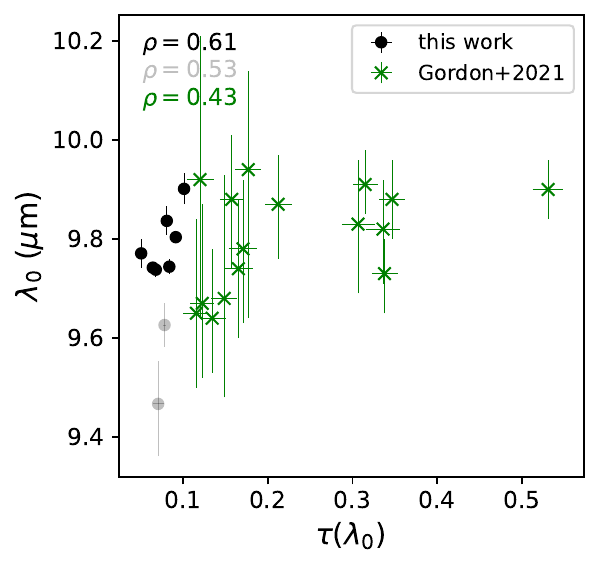}
\caption{Peak wavelength, $\lambda_0$, vs. peak optical depth, $\tau(\lambda_0)$, compared to literature measurements from \cite{2021ApJ...916...33G} (in green). The gray data points correspond to the two sightlines with a noisy MIRI spectrum. The Spearman's rank correlation coefficients are shown in the corner (in black for the black data points, in gray for all the data points in this work, and in green for all the data points in this plot).
 \label{fig:sil_params_lit}}
\end{figure}

In addition, there appears to be a moderate correlation between $\tau(\lambda_0)$ ($\rho=0.61$) (or integrated area ($\rho=0.71$)), and the peak wavelength, $\lambda_0$, of the feature. Stronger features seem to peak at slightly longer wavelengths, but the scatter is substantial. As explained in Sec.~\ref{sec:fit_sil}, \cite{2021ApJ...916...33G} used a modified Drude profile to fit this feature. Therefore, the only parameters that are directly comparable with ours are the peak wavelength and the peak optical depth\footnote{\cite{2021ApJ...916...33G} fit the continuum extinction and both silicate features simultaneously from the extinction curve, whereas we fit only the 10\,\micron\ feature after first normalizing the spectrum around that feature. This different approach can potentially result in a small offset between our peak optical depths, with smaller values in our case.}. When adding their data points to our plot, the (moderate) trend between $\tau(\lambda_0)$ and $\lambda_0$ becomes weaker ($\rho=0.43$), as can be seen in Fig.~\ref{fig:sil_params_lit}.

\cite{2021ApJ...916...33G} find a very strong anti-correlation between asymmetry and peak wavelength. They suggest that this could mean that the feature can likely be described with fewer parameters (i.e., these properties are not entirely independent).
We also find a strong anti-correlation ($\rho=-0.89$) between $\alpha$ and $\lambda_0$, with more asymmetric features peaking at shorter wavelengths. We want to caution here about a possible degeneracy between these two parameters. For some sightlines, a similarly good fit to the feature could be obtained with a slightly longer peak wavelength and a slightly smaller $\alpha$, which could partially explain the anti-correlation seen between these two parameters.
We found that $\alpha$ does not correlate with $\tau(\lambda_0)$ or with the integrated area of the feature.

\cite{2021ApJ...916...33G} found a correlation between the peak wavelength and the width of the feature. \cite{2024ApJ...963..120S} also reported a weak correlation between peak wavelength and FWHM, but their scatter is substantial.
As mentioned before, the FWHM of the feature does not vary significantly between different MEAD sightlines, with a median value of 1.85\,\micron, and a standard deviation of 0.08\,\micron. No correlations were found between the FWHM and other feature properties.

\subsection{Silicate feature vs. extinction}
\subsubsection{A(V), A(1500\,\AA) and R(V)}
\begin{figure*}[ht]
 \centering
  \subfloat{\includegraphics[height=0.55\textwidth]{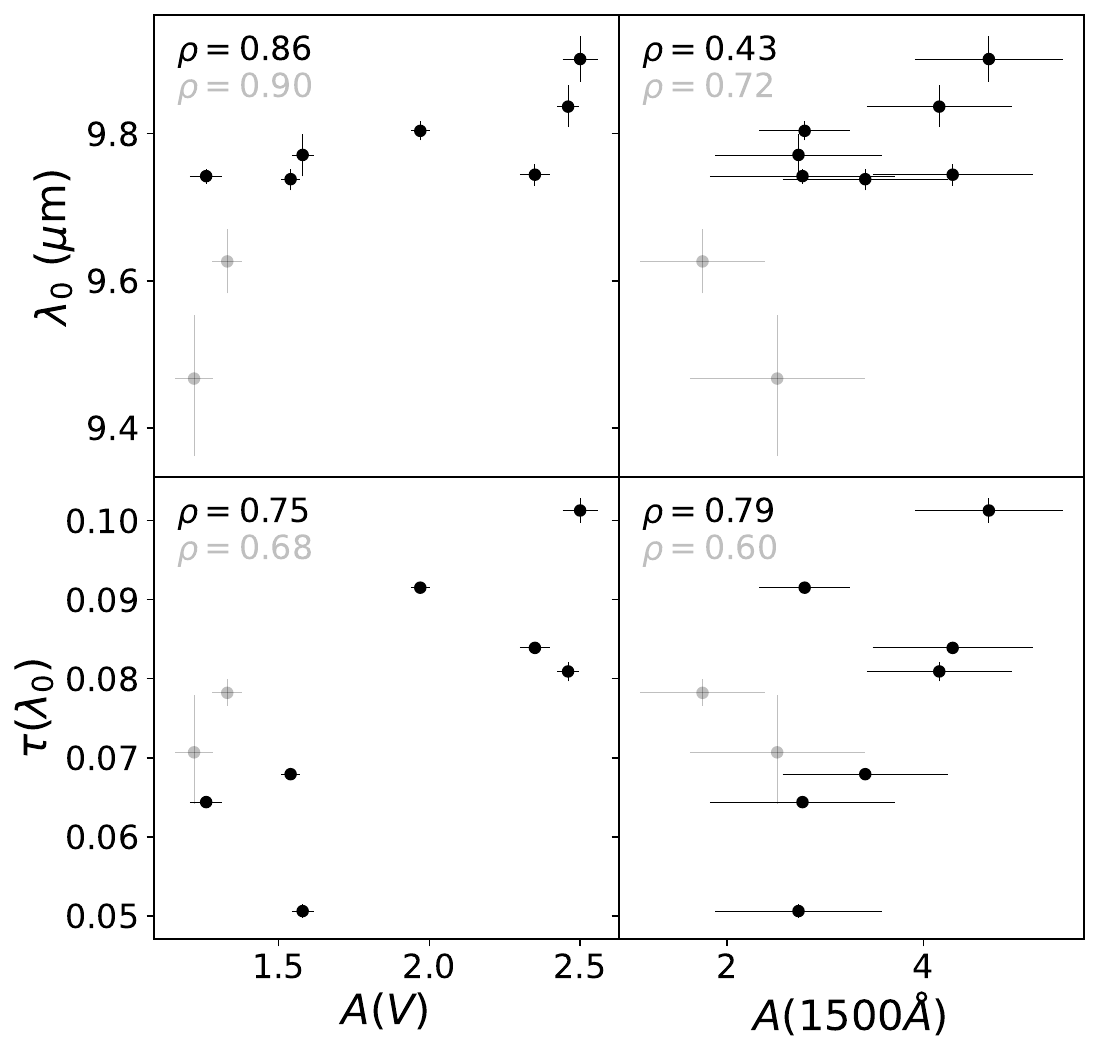}}
  \subfloat{\includegraphics[height=0.55\textwidth]{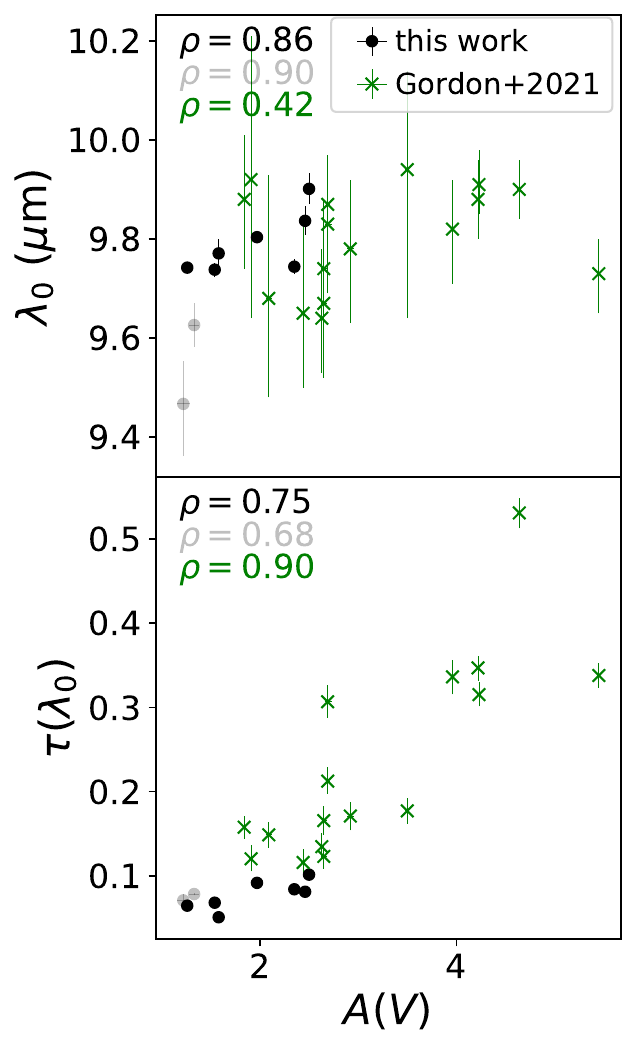}}
   \caption{Silicate feature properties compared to $A(V)$ and $A(1500\,\text{\AA})$, with literature measurements from \cite{2021ApJ...916...33G} added in the right panels (in green).}
  \label{fig:feat_AV}
\end{figure*}

\cite{2024ApJ...963..120S} reported a strong correlation between peak optical depth of the silicate feature and the V-band extinction, $A(V)$, for their sample of high $A(V)$ sightlines. Fig.~\ref{fig:feat_AV} (bottom left panel) shows that also in our sample the peak optical depth  of the feature correlates ($\rho=0.75$) with $A(V)$ (taken from \cite{2009ApJ...705.1320G}). The same is true for the integrated area ($\rho=0.79$). When adding the measurements from \cite{2021ApJ...916...33G}, which extends the sample to larger $A(V)$s, the correlation between $\tau(\lambda_0)$ and $A(V)$ becomes much stronger ($\rho=0.90$) (see bottom right panel in Fig.~\ref{fig:feat_AV}). This correlation is not surprising as a larger $A(V)$ is generally consistent with more dust along the line of sight, and hence a stronger extinction feature. Nevertheless, it confirms that silicate grains likely also contribute to the optical extinction, or, at a minimum, that the amount of silicate grains scales proportionally with the amount of grains that cause the continuum extinction at optical wavelengths. At the same time, the correlation is not perfect, indicating that silicates are not solely responsible for the optical extinction. This is consistent with the general picture that carbonaceous grains contribute significantly to the optical extinction as well \citep[see e.g., sec~3.6.2 in][]{2022dge..book.....W}. We also find a strong correlation between $\lambda_0$ and $A(V)$ ($\rho=0.86$) (Fig.~\ref{fig:feat_AV}, top left). However, this trend becomes much weaker ($\rho=0.42$) when adding the data from \cite{2021ApJ...916...33G} (Fig.~\ref{fig:feat_AV}, top right). We argue that this could be a secondary correlation, as a result of the correlations between $\lambda_0$ and $\tau(\lambda_0)$, and between $\tau(\lambda_0)$ and $A(V)$.

Certain dust grain models (such as \cite{2001ApJ...548..296W}, \cite{2013A&A...558A..62J} and some of the \cite{2004ApJS..152..211Z} models) show that the extinction around 1500\,\AA\ is dominated by silicate grains.
We calculated $A(1500\,\text{\AA})$ for each of our sightlines by evaluating the FM90 extinction curve model \citep[][see next subsection]{1990ApJS...72..163F} at 1500\,\AA, with the \texttt{dust\_extinction} package \citep{Gordon2024}, using the UV extinction curve parameters (as well as $E(B-V)$ and $A(V)$) for our sightlines obtained by \cite{2009ApJ...705.1320G}. The uncertainty (standard deviation) on $A(1500\,\text{\AA})$ for every sightline was determined by evaluating the FM90 model 1000 times with UV extinction parameter combinations drawn randomly from a multi-variate Gaussian distribution with as standard deviations the uncertainties on these individual parameters (as reported in \cite{2009ApJ...705.1320G}). As shown in Fig.~\ref{fig:feat_AV} (bottom middle panel), the peak optical depth of the feature indeed seems to correlate better ($\rho=0.79$) with $A(1500\,\text{\AA})$ than with $A(V)$\footnote{The correlation coefficient is only slightly larger, however, visually, the correlation looks stronger. We note that the correlation coefficient is only one way to interpret the data, and does not necessarily tell the complete story, especially for a relative small sample like ours. We found that similar looking patterns sometimes corresponded to quite different coefficients, and we caution that these coefficients can thus be misleading.}, which is consistent with silicate grains dominating the extinction around 1500\,\AA\ as suggested in the aforementioned dust grain models (although the uncertainties on $A(1500\,\text{\AA})$ are substantial).
There is, however, no clear trend between peak wavelength and $A(1500\,\text{\AA})$ ($\rho=0.43$, Fig.~\ref{fig:feat_AV}, top middle).

We note that no correlations were found between the FWHM or the asymmetry of the feature and $A(V)$ or $A(1500\,\text{\AA})$.

Furthermore, there are no (convincing) trends between the silicate feature properties and $R(V)$. This remains the case when adding the data from \cite{2021ApJ...916...33G} for peak wavelength and peak optical depth.

When normalizing the peak optical depth (or integrated area) of the feature by $A(V)$, as a proxy for the total dust content along the line of sight, we found a (weak) anti-correlation with $A(V)$ ($\rho=-0.46$) and with $R(V)$ ($\rho=-0.43$). Sightlines with larger $A(V)$ or larger $R(V)$ seem to have weaker silicate features relative to the total dust extinction. In other words, these sightlines seem to have a relatively smaller silicate contribution compared to other dust grain types such as carbonaceous grains. However, after adding the measurements from \cite{2021ApJ...916...33G} these trends disappeared.
Even sightlines with similar values of $A(V)$ or $R(V)$ have significantly different silicate feature strengths. This, again, indicates that silicates cannot be responsible for all the optical extinction. Different sightlines have relatively higher or lower silicate contributions compared to other dust grain types such as carbonaceous grains, and this variation does not scale with the amount of dust (probed by $A(V)$) or with the average dust grain size (probed by $R(V)$).

In our sample, we measured an $A(V)/\tau(\lambda_0)$ between $\sim$17 and $\sim$31, with an average of 23 and a standard deviation of 5. In the literature, a range of $A(V)/\tau(\lambda_0)$ has been reported. \cite{2010EP&S...62...63G}, e.g., collected different literature values for the local diffuse ISM between 16.7 and 19.2, with an average of 18.2. More recently, \cite{2021ApJ...916...33G} found  an average $A(V)/\tau(\lambda_0) \approx13$, but with a large variation between their sightlines (between $\sim$9 and $\sim25$), and \cite{2024ApJ...963..120S} found an average value around 18.
Our average value is somewhat larger than most literature values. This could indicate that our optical depths are somewhat underestimated due to the local continuum normalization (as mentioned above), or that our low $A(V)$, purely diffuse, sample probes different ISM conditions compared to those studied in the literature, with for example a larger relative contribution of carbonaceous grains. Further investigation with a larger sample of low $A(V)$ sightlines is necessary to better understand these measured variations in $A(V)/\tau(\lambda_0)$.

\subsubsection{UV extinction curve properties}
Correlations between different extinction features could reveal additional clues on the carriers of those features. 
UV extinction curves have been measured for the MEAD sample by \cite{2009ApJ...705.1320G}. A UV extinction curve can generally be described by the FM90 parameters \citep{1990ApJS...72..163F}. The FM90 formulation uses a combination of a line for the continuum UV extinction, a Drude profile for the 2175\,\AA\ bump, and a quadratic function for the far-UV (FUV) rise. We did not find any significant correlations between the silicate extinction feature properties and the UV extinction properties (except for $A(1500\,\text{\AA})$, see previous subsection) including the UV continuum slope, UV bump amplitude, UV bump integrated area and FUV rise. This indicates that these UV extinction features (bump and FUV rise) and the silicate feature are not caused by the same (type of) grains. \cite{2021ApJ...916...33G} also did not find a correlation between the silicate feature and the UV bump in their sample of sightlines. This is consistent with the general picture that the UV bump and FUV rise are caused by carbonaceous grains \citep[e.g.,][]{1965ApJ...142.1681S, 2003ARA&A..41..241D,2010ApJ...712L..16S}.

\subsection{Silicate feature vs. hydrogen columns}

\begin{figure*}[ht]
\centering
\includegraphics[width=\textwidth]{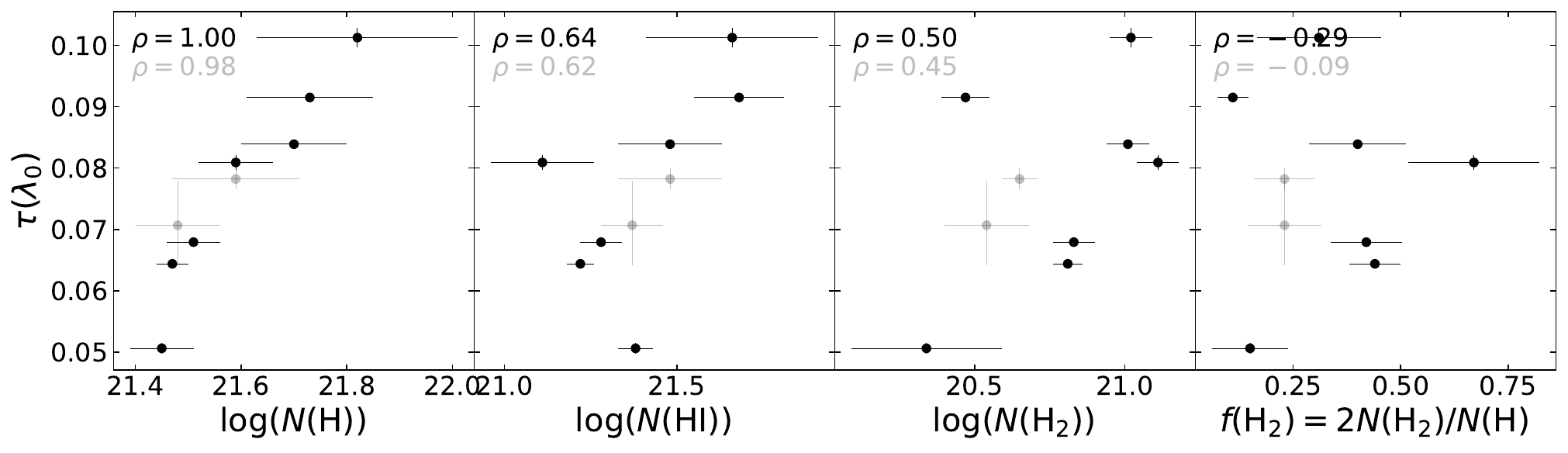}
\caption{Silicate feature peak optical depth compared to hydrogen column densities and molecular hydrogen fraction.
 \label{fig:feat_H}}
\end{figure*}

As gas and dust are generally well mixed in the ISM \citep[e.g.,][]{2022dge..book.....W}, we expect a correlation between dust extinction and the total hydrogen column density along the line of sight \citep[see e.g.,][]{1978ApJ...224..132B}. Furthermore, atomic and molecular hydrogen trace different phases of the gas in the ISM, and dust properties could vary between these different ISM conditions. Fig.~\ref{fig:feat_H} explores the relationships between the silicate feature peak optical depth and hydrogen column densities, taken from \cite{2023ApJ...944...33V}. We find a very strong correlation between the peak optical depth ($\rho=1$, left panel) (or integrated area, $\rho=0.96$) of the feature and the total hydrogen column density, $N$(H). This indicates that there is more (silicate) dust in sightlines with a larger total hydrogen content, as expected. Interestingly, the correlation is weaker when considering the atomic (\ion{H}{1}, $\rho=0.64$, second panel) and molecular (H\textsubscript{2}, $\rho=0.50$, third panel) hydrogen separately. This suggests that silicate grains are present in both gas phases, and not only in atomic or only in molecular environments, i.e., the silicate dust is gas-phase independent. The molecular hydrogen fraction, $f(\text{H}_2)\ (=2N(\text{H}_2)/N(\text{H})$), is often used as a tracer of the gas cloud density (or the environment) along the line of sight. We did not find a significant correlation between the silicate feature peak optical depth and $f(\text{H}_2)$ ($\rho=-0.29$, right panel). This suggests that the amount of silicate grains does not directly depend on the ISM conditions (as traced by $f(\text{H}_2)$).

There are no significant correlations between the peak wavelength or FWHM of the feature and the hydrogen column densities or molecular hydrogen fraction.

\subsection{Silicate feature vs. elemental columns in dust}\label{sec:cols}
As explained in the Introduction, the goal of the MEAD project is to combine extinction and elemental abundance measurements in the same sightlines in order to put stronger constraints on the dust grain properties. We are in the process of measuring elemental abundances and column densities in the gas and dust in the MEAD sample for C, Si, Mg, Fe, and O, from our obtained HST/STIS spectra. The details and results of those measurements will be presented in a separate dedicated paper (MEAD paper II, Decleir et al., in prep.). Here, for the first time, we compare literature abundance (column density) measurements that are available for the MEAD sightlines with our measured silicate extinction feature, as a proof of concept.

We obtained column densities in the gas, $N(\text{X})_{\text{gas}}$, for Mg, Fe and O from \cite{2023ApJ...952...57R} for a subset of the MEAD sightlines (except for HD203938 which was taken from \cite{2009ApJ...700.1299J}). We calculated the column density of element X in dust as:
\begin{equation}
N(\text{X})_{\text{dust}} = N(\text{H}) [N(\text{X})_{\text{ref}}/N(\text{H})] - N(\text{X})_{\text{gas}}
\end{equation}

\noindent where $N(\text{X})_{\text{ref}}/N(\text{H})$ is a total reference abundance (i.e., gas+dust) for element X (taken to be the solar abundance for that element, listed in \cite{2009ApJ...700.1299J}), and with $N(H)$ for our sightlines  from \cite{2023ApJ...944...33V}.

\begin{figure*}
\centering
\includegraphics[width=0.9\textwidth]{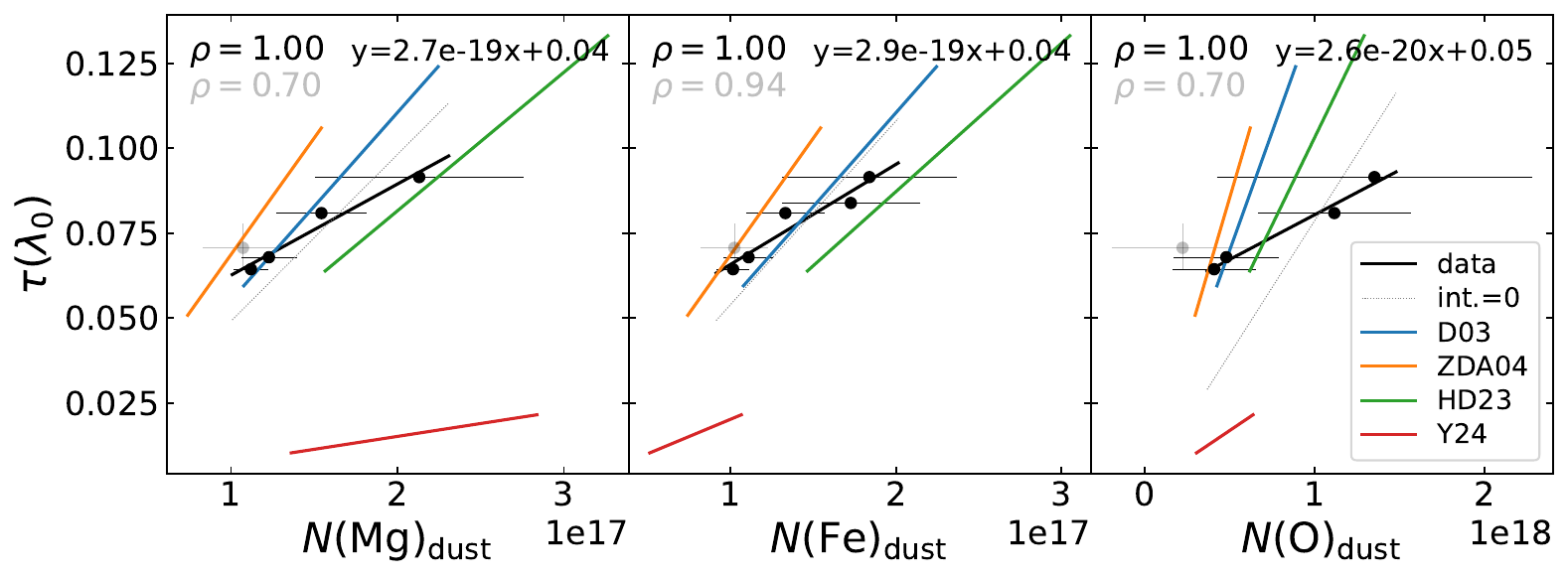}
\caption{Silicate feature peak optical depth compared to column densities of Mg, Fe and O in dust, taken from \cite{2023ApJ...952...57R} (or \cite{2009ApJ...700.1299J}). The black line is a fit to the black data points, with its equation given in the top right corner of each panel. The colored lines correspond to different dust grain models for $A(V)$s between 1.2 and 2.5.
 \label{fig:feat_dcol}}
\end{figure*}

As shown in Fig.~\ref{fig:feat_dcol}, we find strong correlations between the peak optical depth of the silicate feature and the column densities of Mg, Fe and O in dust ($\rho=1$).
It is commonly accepted that Mg- and Fe-rich silicates are responsible for this extinction feature, e.g. olivines such as forsterite (Mg\textsubscript{2}SiO\textsubscript{4}) and fayalite (Fe\textsubscript{2}SiO\textsubscript{4}), and pyroxenes such as enstatite (MgSiO\textsubscript{3}) and ferrosilite (FeSiO\textsubscript{3}) \citep[see e.g. the review by][]{2010ARA&A..48...21H}. The observed, tight correlations between the measured feature strength (from IR spectra) and the independently measured column densities in dust (from UV spectra) are consistent with silicate grains being the carrier of the feature, as previously established. Alternatively, it is also possible that the Mg, Fe and O are locked up in another type of dust grains (such as metal oxides) that form and exist under the same conditions as silicates.

We fit the black data points in Fig.~\ref{fig:feat_dcol} with a line, using the Astropy \texttt{LinearLSQFitter} and the \texttt{Linear1D} model. Interestingly, we find that the fitted line has a positive intercept for all three elements. This means that if we extrapolate this trend to zero column density for Mg, Fe and O in dust, we would still measure a (weak) silicate feature, which is not realistic. To guide the eye, we added a dotted gray line, obtained by forcing the fit to have a zero intercept. The data seem to underestimate the elemental column densities in dust and/or overestimate the strength of the silicate feature. This could be caused by a range of assumptions that were made when fitting the feature (e.g., the continuum level, or the shape of the feature), or when measuring the column densities (e.g., reference solar abundances, absorption line oscillator strengths, or the method used to measure gas column densities), or a combination of the above. It is also possible that we are affected by the relatively small sample size, and that the measured slopes would change if more data would be added. Alternatively, maybe the assumption of a linear relationship between feature strength and elemental column density in dust is not valid, which could be the case if the stoichiometry of the silicates changes significantly in sightlines with lower dust column densities. Furthermore, the elemental column densities in the dust are derived from column densities in the gas, and could probe a somewhat different mix of ISM phases along the line of sight, compared to the direct dust extinction measurement. Finally, a positive intercept could also be explained if the 10\,\micron\ extinction feature is not entirely caused by Mg- and Fe-rich silicates, and some other type of dust grain also contributes to this feature. It is beyond the scope of this work to find the exact reason for this positive intercept, and it is clear that we need more data, especially at lower dust column densities, to better understand these measurements.

\begin{deluxetable*}{l|ccccccr@{ :}c@{: }l}
\tablecaption{Slopes of lines in Fig.~\ref{fig:feat_dcol} for the fitted line to the data points, and for the dust grain models, and ratios between slopes. \label{tab:grmods}}
\tablehead{\colhead{line} & \colhead{$\tau/N$(Mg)} & \colhead{$\tau/N$(Fe)} & \colhead{$\tau/N$(O)} & \colhead{$N$(Mg)/$N$(Fe)} & \colhead{$N$(Mg)/$N$(O)} & \colhead{$N$(Fe)/$N$(O)} & \colhead{Mg :} & \colhead{ Fe :} & \colhead{O}}
\startdata
data & 2.7e-19 & 2.9e-19 & 2.6e-20 & 1.10 & 0.10 & 0.09 & 1.1 & 1 & 11.2 \\
D03 & 5.5e-19 & 5.5e-19 & 1.4e-19 & 1.00 & 0.25 & 0.25 & 1 & 1 & 4 \\
ZDA04 & 6.9e-19 & 6.9e-19 & 1.7e-19 & 1.00 & 0.25 & 0.25 & 1 & 1 & 4 \\
HD23 & 4.1e-19 & 4.4e-19 & 1.0e-19 & 1.07 & 0.25 & 0.24 & 1.1 & 1 & 4.2 \\
Y24 & 7.6e-20 & 2.0e-19 & 3.4e-20 & 2.67 & 0.44 & 0.17 & 2.7 & 1 & 6
\enddata
\end{deluxetable*}

We added dust grain models to these plots: \cite{2003ARA&A..41..241D,2003ApJ...598.1017D} (D03), \cite{2004ApJS..152..211Z} (ZDA04), \cite{2023ApJ...948...55H} (HD23), and \cite{2024A&A...684A..34Y} (Y24). For each model, we obtained the peak optical depth of the silicate feature by fitting the model in the same way as we fit the data. I.e., we used the model extinction curve to mimic an extinguished stellar spectrum (assuming a flat stellar spectrum of 1 and an $A(V)=1$) with the \texttt{dust\_extinction} package \citep{Gordon2024}. We then fit the continuum with a line, and the feature with a skewed Gaussian, exactly as we did for the data. This gives us, for every model, a value for the peak optical depth of the silicate feature per $A(V)$. We also obtained the abundances for Mg, Fe and O in dust assumed in each grain model, and converted them into dust column densities per $A(V)$. The colored lines in Fig.~\ref{fig:feat_dcol} show, for every model, how the silicate feature and dust column densities scale for $A(V)$s between 1.2 and 2.5 (which is the range covered by our sample). We assume here that all Mg, Fe and O in the models is incorporated in the silicate grains (i.e., the intercept of the line is zero). The slopes of these lines are given in Table~\ref{tab:grmods} for the data and the models. We find substantial differences between the slopes of the different models. Those of HD23 and Y24 are closest to the data, however, the silicate feature of Y24 is much weaker.

Dividing the slopes of the fitted black lines in Fig.~\ref{fig:feat_dcol} results in a measurement of the average stoichiometry of the silicates observed in the MEAD sightlines. For example, dividing $\tau/N$(Fe) by $\tau/N$(Mg), gives the ratio $N$(Mg)/$N$(Fe) (also listed in Table~\ref{tab:grmods}). We find an average stoichiometry of Mg:Fe:O of 1.1:1:11.2, which is substantially different from the assumptions in the dust grain models. Our sightlines seem to have much more O compared to Mg or Fe in the dust than is assumed in the dust grain models. This might be due to a different observed silicate stoichiometry in the MEAD sightlines than assumed in the models. Differences in stoichiometry are expected to cause changes in the peak wavelength and width of the silicate feature \citep[e.g.,][]{2010ARA&A..48...21H}. However, we did not find any clear correlations between the peak wavelength or FWHM of the feature and the elemental column densities in dust or the ratios thereof, but we note that our sample might be too small to definitively conclude this. In addition, the feature properties can also be influenced by other grain properties such as size, shape or crystallinity, and it is difficult to disentangle these effects from differences in stoichiometry.
Alternatively, it is possible that not all O in the dust is locked up in silicate grains, but is present in some other dust material that is strongly correlated with the silicate dust. For example, if there is indeed water ice (or trapped solid-state water) on the surface of silicate grains, as suggested in Sec.~\ref{sec:ice}, some of the extra O can be accounted for and is not needed for the silicate grain itself. We note that to determine the exact stoichiometry and to constrain the mineral components (olivines/pyroxenes) of the silicates, we also need measurements of the Si column densities in gas, which are currently not available.

A more in-depth investigating of the correlations between elemental column densities in dust and feature properties will be done in a future paper, after finalizing our new elemental column density measurements, including for Si, for all MEAD sightlines (MEAD papers II and III, Decleir et al., in prep.).

\section{Summary and conclusions}\label{sec:conclusions}
The goal of MEAD (Measuring Extinction and Abundances of Dust) is to constrain the dust grain properties in the diffuse ISM of the Milky Way, by combining dust extinction and elemental abundance measurements for the same sample of sightlines. In this first MEAD paper, we studied the extinction features observed in our JWST NIRCam grism and MIRI MRS spectra for nine Milky Way sightlines. In all sightlines, we observed a strong silicate feature around 10\,\micron\ and fit its profile with a skewed Gaussian.

The main conclusions of this work are:
\begin{itemize}
    \item For the first time, we found a strong correlation between the silicate feature strength and the column densities of Mg, Fe and O in dust. This correlation is consistent with the generally accepted attribution of this feature to Mg- and Fe-rich silicate grains. We derived an average stoichiometry of the silicate grains observed in our sample of Mg:Fe:O = 1.1:1:11.2.

    \item The strength of the silicate feature correlates well with $A(V)$, showing that silicates significantly contribute to the optical extinction. However, the feature correlates even better with $A(\text{1500\,\AA})$, indicating that the extinction around 1500\AA\ is dominated by silicates, as suggested by some dust grain models \citep{2001ApJ...548..296W,2013A&A...558A..62J,2004ApJS..152..211Z}.
    
    \item We found variations in the silicate feature peak wavelength, confirming previous studies \citep[e.g.,][]{2021ApJ...916...33G, 2024ApJ...963..120S}. This indicates that different sightlines contain different types of silicate grains.
   
    \item We did not observe any clear correlations between the silicate feature properties and the 2175\,\AA\ bump properties, confirming previous work \citep[e.g.,][]{2021ApJ...916...33G}. This is consistent with the generally accepted picture that these features are not caused by the same type of dust grains. 
    \item We observed features around 3.4 and (tentatively) around 6.2\,\micron\ in the average spectrum of our sample, with a maximum $\tau(3.4\,\micron)/A(V) \approx 0.0035 \pm 0.0009$ and $\tau(6.2\,\micron)/A(V) \approx 0.0032 \pm 0.0006$. These are likely caused by aliphatic and aromatic/olefinic hydrocarbons, respectively, which are expected in the diffuse ISM. If confirmed, to the best of our knowledge, this is the first detection of hydrocarbons in sightlines with an $A(V)\leq2.5$. 
   
    \item We detected a 3\,\micron\ feature in the sightline HD073882 ($A(V)\approx2.5$), with a peak $\tau/A(V) \sim 0.020 \pm0.001$, likely caused by water ice. We also tentatively detected a weak 3\,\micron\ feature in the average of the rest of the sample with a peak $\tau/A(V) \approx 0.0065 \pm 0.0009$. If confirmed, this is the first detection of ice (or trapped solid-state water) in sightlines with an $A(V)\leq2.5$. 
\end{itemize}

In future work (MEAD paper III, Decleir et al., in prep.), we will derive complete NIR--MIR extinction curves for the MEAD sample using stellar atmosphere models and analyze the continuum extinction as well as the extinction features in more detail.

The NIRCam grism and MIRI MRS spectra for the MEAD sightlines are electronically available\footnote{\url{https://doi.org/10.5281/zenodo.14286122}} \citep{decleir_2024_14286122}. The code developed for
the analysis, plots and tables in this work is available as part of the \texttt{mead} package on GitHub\footnote{\url{https://github.com/mdecleir/mead/releases/tag/v1.0.0}} \citep{marjorie_decleir_2024_14291651}.

\vspace{5mm}
\facilities{JWST (NIRCam grism, MIRI MRS), IUE, FUSE}

\begin{acknowledgments}

MD and SZ acknowledge support from the Research Fellowship Program of the European Space Agency (ESA). BG acknowledges support of TÜBİTAK 2219 program.

This work is based on observations made with the NASA/ESA/CSA James Webb Space Telescope. The data were obtained from the Mikulski Archive for Space Telescopes at the Space Telescope Science Institute, which is operated by the Association of Universities for Research in Astronomy, Inc., under NASA contract NAS 5-03127 for JWST. These observations are associated with program \#2459. Support for program \#2459 was provided by NASA through a grant from the Space Telescope Science Institute, which is operated by the Association of Universities for Research in Astronomy, Inc., under NASA contract NAS 5-03127.

This research has made use of the SIMBAD database, operated at CDS, Strasbourg, France. This research has made use of the NASA/IPAC Infrared Science Archive, which is funded by the National Aeronautics and Space Administration and operated by the California Institute of Technology. This publication makes use of data products from the Two Micron All Sky Survey, which is a joint project of the University of Massachusetts and the Infrared Processing and Analysis Center/California Institute of Technology, funded by the National Aeronautics and Space Administration and the National Science Foundation. This publication makes use of data products from the Wide-field Infrared Survey Explorer, which is a joint project of the University of California, Los Angeles, and the Jet Propulsion Laboratory/California Institute of Technology, funded by the National Aeronautics and Space Administration. 

\end{acknowledgments}

\software{Astropy \citep{astropy:2013,astropy:2018,astropy:2022}, SciPy \citep{2020SciPy-NMeth}, NumPy \citep{harris2020array}, IDL, Emcee \citep{2013PASP..125..306F}, Matplotlib \citep{Hunter:2007}, dust\_extinction \citep{Gordon2024}, mead \citep{marjorie_decleir_2024_14291651}, nircam\_grism \citep{2023ApJ...953...53S}}

\bibliography{references}{}
\bibliographystyle{aasjournal}

\end{document}